\documentclass{raa}

\usepackage{graphicx,times}             

\begin{document}

   \title{{R}elations {between} stellar mass and electron
  temperature-based metallicity {for} star-forming galaxies in a
  wide mass range $^*$
\footnotetext{\small $*$ Supported by the National Natural Science
Foundation of China.} }

  \volnopage{{\bf 2014} Vol.\ {\bf 14} No. {\bf 7}, 875--890~~
 {\small doi: 10.1088/1674--4527/14/7/010}}
   \setcounter{page}{875}

\author{Wei-Bin Shi\inst{1,2}, Yan-Chun Liang\inst{2}, 
   Xu Shao\inst{2,3}, Xiao-Wei Liu\inst{4}, Gang Zhao\inst{1,2},
   \\[1mm]
  Francois Hammer\inst{5}, Yong Zhang\inst{6}, Hector Flores\inst{5},
  Gui-Ping Ruan\inst{1,2} and Li Zhou\inst{1,2,3}
   \vs}

   \institute{Shandong Provincial Key Laboratory of Optical Astronomy
     and Solar-Terrestrial Environment, School of Space Science and
     Physics, Shandong University, 
      Weihai 264209, China \\
     \and
     Key Laboratory of Optical Astronomy, National
     Astronomical Observatories,
Chinese Academy of Sciences, Beijing 100012, China; {\it ycliang@bao.ac.cn} \\
     \and
     University of Chinese Academy of Sciences, Beijing 100049,  China \\
     \and
     Kavli Institute for Astronomy and Astrophysics,
     Peking University, Beijing 100871, China \\
       \and
  GEPI, Observatoire de Paris-Meudon, Meudon 92195, France \\
       \and
  Department of Physics, University of Hong Kong, Hong Kong , China\\
\vs\no
   {\small Received 2014 February 18; accepted 2014 April 30}}

   \abstract{   We select 947 star-forming galaxies from SDSS-DR7 with
     [O~{\sc iii}]$\lambda$4363 emission lines
     detected at a signal-to-noise {ratio }larger than 5$\sigma$.
     Their electron temperatures and direct oxygen abundances are {then }determined. {W}e
     compare the results from different methods. $t_2${, the} electron
     temperature in {the }low ionization region{,} estimated from $t_3${, that} in {the }high ionization region{,} {is} compared {using}
three analysis
     relations between $t_2-t_3${. These} show obvious differences, which
     result in some different ionic oxygen abundances. The results of
     $t_3$, $t_2$, {$\rm O^{++}$/$\rm H^+$} and {$\rm O^{+}$/$\rm H^+$} derived by using
     methods from IRAF and literature are also compared. The ionic
     abundances $\rm O^{++}$/$\rm H^+$ {are} higher than $\rm O^{+}$/$\rm H^+$ for most
     cases.  The{ different} oxygen abundances derived from $T_{\rm e}$ and the strong-line ratios show {a }clear discrepancy, which is more obvious
     following increasing stellar mass and strong-line ratio
     $R_{23}$. The sample{ of} galaxies from SDSS {with}
     detected [O~{\sc iii}]$\lambda$4363 have lower metallicites and
     higher {star formation rates}, {so} they may not be typical representatives of the
     whole{ population of} galaxies. Adopting data objects from
     {Andrews \& Martini}, {Liang et al.} and {Lee et al.} data, we derive new relations of stellar mass and metallicity for star-forming
     galaxies in a much wider stellar mass range: from $10^6\,M_\odot$
     to $10^{11}\,M_\odot$.
   }

   \authorrunning{W. B. Shi et al.}
   \titlerunning{Relations of Stellar Mass and Metallicity of Galaxies }

  \vs \maketitle

\section{Introduction}

Stellar mass and metallicity are two of the most fundamental
physical properties of galaxies. Stellar mass reflects the amount
of gas locked up in stars, {while} metallicity {represents} the
gas reprocessed by stars and any exchange of gas between the
galaxy and its environment (Tremonti et al. 2004).

To estimate accurate{ values of} metallicit{y} for the ionized gas
in galaxies, the electron temperature ($T_{\rm e}$) in the gas is
generally{ required}, which is usually obtained from the ratio of
auroral to nebular line intensities, such as [O~{\sc iii}]$\lambda
\lambda$4959, 5007/[O~{\sc iii}]$\lambda$4363. This is generally
known as the “direct $T_{\rm e}$-method” since $T_{\rm e}$
{can} be directly inferred from observed line ratios. However, it
is well known that this procedure is difficult to carry out for
metal-rich galaxies since, as the metallicity increases, the
electron temperature decreases (as the cooling is via metal
lines), and the auroral lines eventually become too faint to
measure. Instead, other strong nebular line ratios {are required}
to estimate the oxygen abundances of metal-rich galaxies ($12 +
\log \rm (O/H) > 8.5$) (Pagel et al. 1979; Tremonti et al. 2004;
Liang et al. 2006; Kewley \& Dopita 2002).

{F}or estimating electron temperatures $T_{\rm e}$, and then the
$T_{\rm e}$-derived oxygen abundances for galaxies, several methods
have been given in {the }literature. In {the }first
part of this work, we compare these different
methods based on a large sample of good quality data from {the Sloan Digital Sky Survey (}SDSS{)}.
{T}he SDSS survey provides a large
sample of galaxies {whose}
[O~{\sc iii}]$\lambda$4363 fluxes{ have been measured}, and
{whose} $T_{\rm e}$-based metallicities could be
{reliably }estimated.

The stellar mass and metallic{i}ty{ relation} (MZR) of galaxies is
a fundamental relation{ship} {that} {represents} the evolutionary
history and the present properties of galaxies. Generally, the
metallicities and stellar masses of galaxies increase with their
evolutionary processes. Therefore, usually more massive galaxies
are more metal rich (Tremonti et al. 2004; Liang et al. 2004,
2007; Kewley \& Ellison 2008, and the references therein).
\cite{tre04} gave the origin of the mass metallicity relation from
{53\,000} star-forming galaxies in the SDSS. Several works have
been trying to obtain the MZR from $T_{\rm e}$-based oxygen
abundances.

However, for metal-rich galaxies, it is difficult to obtain their
[O~{\sc iii}]4363 and then $T_{\rm e}$-based oxygen abundances{,
s}o the common method{ used} is {to rely on} strong emission line
ratios ({the }so called strong-line method) to estimate their
metallicities. {S}ome researchers {have also }tried to stack the
spectra of multipl{e} galaxies in given bins of stellar masses,
and then derive their [O~{\sc ii}]$\lambda\lambda$7320, 7330 and
$T_{\rm e}$-based oxygen abundances (Liang et al. 2007; Andrews \&
Martini 2013). However, the range of stellar masses{ is} generally
not wide. Therefore, in the second part of this work, we gather
{data} from literature to build a sample {with a} wide{ range of}
stellar mass from 10$^6~M_{\odot}$ to 10$^{11}~M_{\odot}$. {These
data} have been{ used to} obtain $T_{\rm e}$ and $T_{\rm e}$-based
oxygen abundances. Then we derive a relation of stellar mass and
metallicity for galaxies in this wide range of stellar masses.
These sample galaxies are taken from \cite{lia07}, \cite{and13}
and \cite{lee06}. \cite{lee06} gave the mass-metallicity relation
of galaxies at lower stellar mass. \cite{lia07} and \cite{and13}
worked on more metal-rich galaxies from SDSS.

We always keep in mind the difference between the oxygen
abundances obtained from {the }electronic temperature method
($T_{\rm e}$-method) and the strong line ratio method (so-called
$R_{23}$). This {issue }is still in debate (Kennicutt et al. 2003;
Liang et al. 2007; Stasinska 2005). We compare them by using the
good quality data here.

This paper is organized as follows. The sample selection criteria
are described in Section~2. The determinations of $T_{\rm
e}$ from line ratios and oxygen abundances from $T_{\rm e}$ are
presented and compared in Section~3 in detail. In Section~4, we fit
the relations of stellar mass and metallicity in a wide mass range
from the $T_{\rm e}$-based oxygen abundances. We
summar{ize} this work in Section~5.

\section{Sample selection}

We select 947 star-forming galaxies from the SDSS-{Data Release 7
(}DR7{)} main galaxy sample for which their [O~{\sc
iii}]$\lambda$4363 emission lines{ can be measured} at a
signal-to-noise{ ratio} above 5$\sigma$. This auroral line can
help us to {reliably }derive the electron temperature $T_{\rm e}$,
and then {have a} direct{ measure of} oxygen abundances from
$T_{\rm e}$. {W}e{ also} require that these galaxies have stellar
mass estimates. These sample galaxies are selected as follows and
their numbers during the selection process are shown
in Table~\ref{tab1}. 

We download {927\,552} galaxies from the SDSS-DR7
(Adelman-McCarthy et al 2006). Then we select the galaxies with
redshifts 0.03$<z<$0.25 to ensure {we} cover{ the range} from
[O~{\sc ii}] to H$\alpha$ and [S~{\sc ii}] emission lines.
\cite{tre04} also discussed the weak effect of aperture on
estimated metallicities of the sample galaxies with $0.03 < z <
0.25$ and this was further discussed by \cite{kew05}. Then
{747\,970} galaxies remain {after} this criteri{on}{ is applied}.

Following the works of \cite{yin07}, \cite{lia06} and \cite{tre04},
we further select objects displaying [O~{\sc
  ii}]$\lambda\lambda$3726, 3729, [O~{\sc iii}]$\lambda$5007, H$\beta$,
H$\alpha$, [N~{\sc ii}]$\lambda6583${ and} [S~{\sc
  ii}]$\lambda\lambda6717,6731$ emission lines, and H$\beta$,
H$\alpha$, [N~{\sc ii}]{ and} [S~{\sc ii}] emission lines with adequate
{signal to noise (}S/N{)} ($>$5\,$\sigma$). There are {274\,197} galaxies
matching the criteria.

Since we will use the $T_{\rm e}$-method to derive O/H abundances {for}
the sample{ of} galaxies (see Sect.\,3 for details), we
select the samples with{ the} [O~{\sc iii}]$\lambda$4363 emission{ }line
detected at a{n} S/N ratio greater than 5$\sigma$. Then 1843
galaxies (from {274\,197}) are selected from
this criteri{on}.

\begin{figure}

\vs \centering
\includegraphics[angle=-90,width=8.8cm]{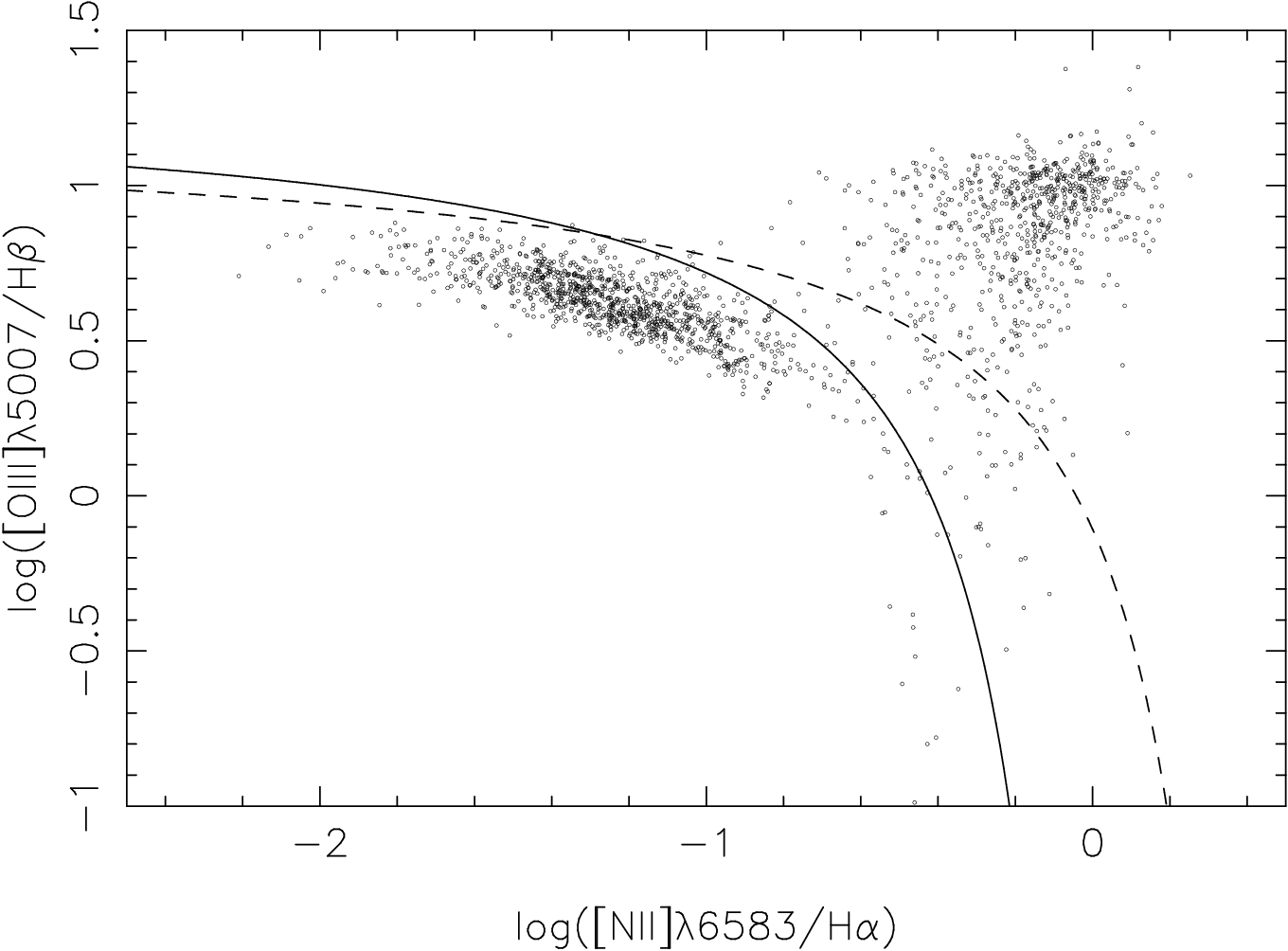}

\caption {\baselineskip 3.6mm A BPT (Baldwin et al.1981) diagram
in which we plot the
  emission-line flux ratio [O~{\sc iii}]/H$\beta$ versus the ratio
  [N~{\sc ii}]/H$\alpha$ for the galaxies in our sample. The solid
  curve shows the demarcation between star-forming galaxies ({\it
  {lower left}}) and {active galactic nuclei} ({\it {upper right}}) defined by \cite{kau03}. {T}he
  dashed line {comes }from \cite{kew01} for reference. Our sample includes
  1072 star-formi{n}g galaxies according to this definition. }
\label{fig.bpt}
\end{figure}

We select 1072 ``star-forming galaxies'' from the 1843
galaxies based on the BPT diagram, which have been identified
following the selection criteria of the traditional line diagnostic
diagram [N~{\sc
  ii}]/H$\alpha$ vs. [O~{\sc iii}]/H$\beta$
(Baldwin et al. 1981; Veilleux \& Osterbrock 1987; Kewley et al.
2001; Kauffmann et al. 2003). These objects are plotted on {the
}BPT diagram as shown in Figure~\ref{fig.bpt}.

To compare our sample galaxies with the whole SDSS main galaxy
sample in{ terms of} their relation {between }star formation rate
 (SFR) {and} stellar
masses as shown in Section~4.1, we also select {211\,725}
``star-forming galaxies'' from
{274\,197} galaxies following {the }BPT diagram
for those objects without [O~{\sc iii}]$\lambda$4363 measurements.
This is taken as the compar{ison} sample.

{Finally}, we obtain 947 star-forming galaxies (see
Table~\ref{tab2} 
  for examples) with metallicity
estimates (from {the }strong-line method) and stellar mass
estimates by{ the} MPA/JHU group. {A} table {listing the} {entire
sample of} galaxies is also provided in {the} electronic version{
of the article}. The {comparison} sample {consists of} {193\,468}
galaxies with measured stellar masses {derived }from the{
compilation of} {211\,725} galaxies mentioned above.


\begin{table*}
\centering \begin{minipage}{80mm}\caption{  List of Detailed
Criteria and Galaxy Number{s} \label{tab1}}\end{minipage}

\fns\tabcolsep 3mm
\begin{tabular}{lc} \hline \noalign{\smallskip} Criteria &
{N}umber{ of Galaxies}  \\\noalign{\smallskip}\hline
\noalign{\smallskip}

{Original }download{ed data set}                           & 927\,552      \\
0.03$<z<$0.25                    & 747\,970      \\
S/N$>$0, S/N$>$5$\sigma$           & 274\,197      \\
{[O~{\sc iii}]} $\lambda 4363>5\sigma$  & 1843        \\
{S}tar forming (BPT)                 & 1072        \\
{S}tellar mass $>$0                  & 947         \\
\noalign{\smallskip} \hline
\end{tabular}
\end{table*}

\begin{table*}
\centering

\begin{minipage}{85mm}

\caption{\baselineskip 3.6mm  An Example Listing of the Galaxies
in Our Sample \label{tab2}}\end{minipage}

\scriptsize \tabcolsep 0.8mm
\begin{tabular}{lcccccccc}
\hline\noalign{\smallskip}
  Num &  Plate-MJD-FiberID &  R{A}-Dec  &  $t_3$ (10$^{4}$K) &
   $t_2$ (10$^{4}$K) &12+(O$^{++}$/H$^+$) &12+(O$^+$/H$^+$) &12+(O/H)$_{\rm Te}$ & 12+(O/H)$_{\rm Bay}$ \\
  (1)  & (2) & (3) & (4)  & (5)   & (6)  &
(7)  & (8) & (9)\\
\noalign{\smallskip} \hline\noalign{\smallskip}
  1& 266-51630-407&145.891$+$   1.117&1.140$\pm$0.023&1.098$\pm$0.016&7.877$\pm$0.029&7.862$\pm$0.027&8.138$\pm$0.028& 8.666\\
  2& 267-51608-421&147.597$+$   0.708&1.313$\pm$0.099&1.219$\pm$0.069&8.041$\pm$0.110&7.422$\pm$0.107&8.053$\pm$0.109& 8.787\\
  3& 270-51909-617&153.629$+$   0.799&1.202$\pm$0.043&1.142$\pm$0.030&8.086$\pm$0.051&7.563$\pm$0.048&8.125$\pm$0.051& 8.396\\
  4& 276-51909-490&163.427$+$   0.163&1.238$\pm$0.022&1.167$\pm$0.015&7.950$\pm$0.024&7.777$\pm$0.023&8.127$\pm$0.023& 8.249\\
  5& 277-51908-451&165.318$+$   0.804&1.525$\pm$0.123&1.367$\pm$0.086&7.968$\pm$0.105&6.983$\pm$0.109&7.913$\pm$0.105& 8.000\\
  6& 280-51612-192&170.438$-$   0.023&1.198$\pm$0.064&1.138$\pm$0.045&8.148$\pm$0.079&7.620$\pm$0.075&8.185$\pm$0.078& 8.171\\
  7& 281-51614-129&172.003$-$   1.127&1.189$\pm$0.045&1.132$\pm$0.032&8.096$\pm$0.054&7.667$\pm$0.051&8.164$\pm$0.054& 8.047\\
  8& 282-51630-546&174.266$+$   0.471&1.162$\pm$0.034&1.113$\pm$0.024&8.234$\pm$0.042&7.639$\pm$0.039&8.253$\pm$0.042& 8.191\\
  9& 282-51658-543&174.266$+$   0.471&1.245$\pm$0.064&1.171$\pm$0.045&8.140$\pm$0.073&7.579$\pm$0.070&8.168$\pm$0.073& 8.188\\
 10& 283-51959-572&176.706$+$   0.896&1.651$\pm$0.244&1.456$\pm$0.171&7.894$\pm$0.215&6.858$\pm$0.225&7.832$\pm$0.215& 7.976\\
\noalign{\smallskip} \hline\noalign{\smallskip}
\end{tabular}
\fns\parbox{135mm}{\baselineskip 3.6mm Notes: The sequence number,
ID number in SDSS and the coordinates are listed in Cols.~1--3.
The electron temperature $t_3$, $t_2$ from {\sc temden.nebular}
and oxygen abundances $\rm O^{++}$/$\rm H^+$ and O$^+$/$\rm H^+$
  from {\sc ionic.nebular} are listed in Cols.~4--7.
$t_2$ is calculated from $t_3$ with the equation $t_2 = 0.7t_3 +
0.3$. 12+(O/H)$_{\rm Bay}${,} obtained by the MPA/JHU group{,} and
is listed in last column.
  The {entire} table of{ galaxies in our} sample is provided in {the}
  electronic~version.} 
\end{table*}

In Section~3, we will use the{ sample of} 947 galaxies to compare the
different methods to estimate $T_{\rm e}$ and then oxygen abundances.

For the second part of this work presented in Section~4, we will study
the stellar mass and metallicity relations of galaxies, and derive new
relations from galaxies in a wide{ range of} stellar mass{es}. Therefore, we
will also collect data with $T_{\rm e}$ and $T_{\rm e}$-based oxygen abundance
estimates from literature: 30 data points from \cite{and13} and 27
data points from \cite{lia07} for higher stellar mass and higher
metallicities,{ and} 25 data points from \cite{lee06} for lower mass and
lower metallicities. The former two are the stacked spectra of SDSS
galaxies in each of the different stellar mass bins. More details will
be discussed in Section~4.

\section{Calculating electron temperature $T_{\lowercase {\rm e}}$
and $T_{\lowercase {\rm e}}$-based oxygen
  abundances for{ galaxies in the} sample}

\subsection{Comparing Electron Temperatures from Different Methods}

A two-zone model for the temperature structure within the H\,{\sc
ii} region was adopted (also see Yin et al. 2007). In this model,
$T_{\rm e}$([O~{\sc iii}]) is taken to represent the temperature
for high{ }ionization species such as $\rm O^{++}$, while $T_{\rm
e}$([O~{\sc ii}]) is used for low{ }ionization species such as
$\rm O^{+}$.

In principle, the temperature {in} {the }high{ }ionization region
$t_3$ ($=10^{-4}\,T_{\rm e}$ ([O~{\sc iii}])) can be derived from
the emission{ }line ratio of [O~{\sc iii}]$\lambda\lambda$4959,
5007/[O~{\sc
  iii}]$\lambda$4363, while the temperature in the low ionization region $t_2$
($=10^{-4} T_{\rm e}$ ([O~{\sc ii}])) can be estimated from
[O~{\sc
  ii}]$\lambda\lambda$7320, 7330/[O~{\sc ii}]$\lambda$3727 and
[N~{\sc ii}]$\lambda$6548, 6583/[N~{\sc ii}]$\lambda$5755.  But in
the spectra, it is often difficult to detect the weak [O~{\sc ii}]
and [N~{\sc ii}] lines except {in} some H~{\sc ii} regions
(Kennicutt et al. 2003, Bresolin et al. 2004) and the stacked
spectra of galaxies (Liang et al. 2007, Andrews \& Martini 2013).
[O~{\sc iii}]$\lambda$4363 is more commonly detected {in}
metal-poor H~{\sc ii} regions and star{ }forming galaxies. Thus,
the common method is {first} deriving{ temperature} $t_3$ from
[O~{\sc iii}]$\lambda\lambda$4959, 5007/[O~{\sc
  iii}]$\lambda$4363 and then estimating $t_2$ from an
analytical relation between $t_2$ and $t_3$ inferred from
photoionization calculations.

The electron densities {$n_{\rm e}$}{ in the ionized gas of the
galaxies can be calculated at $T_{\rm e}=10\,000$~K from the line
ratios [S~{\sc
  ii}]$\lambda6717$/[S~{\sc ii}]$\lambda6731$ by using the five-level
statistical equilibrium model in the task {\sc temden.nebular}
contained in the {\sc iraf/stsdas} package (de Robertis, Dufour \&
Hunt 1987; Shaw \& Dufour 1995), which uses the latest atomic
data. We adopt this method to calcul{a}te {$n_{\rm e}$}.

\subsubsection{Comparing $t_3$ from different methods}

The electron temperature $t_3$ (in units of $10^4$~K) can be
estimated from the iterative formula given by \cite{izo06}{,}
which was taken from \cite{aller84}
\begin{equation}
t_3 = \frac{1.432}{\log ((\lambda4959 + \lambda5007)/\lambda4363)
- \log C_{\rm T}}\, , \label{eq1}
\end{equation}
where
\begin{equation}
C_{\rm T} = \Big(8.44 - 1.09t_3 + 0.5t_3^2 - 0.08t_3^3\Big)\frac{1
+ 0.0004x_3}{1+0.044x_3}\, , \label{eq2}
\end{equation}
with $x_3 = 10^{-4}n_{\rm e}t_3^{-1/2}$, and {$n_{\rm
e}$} is the electron density in cm$^{-3}$. They used atomic data
from the references listed in \cite{sta05}.

The electron temperature $t_3$ can also be estimated by using the
task {\sc temden.nebular} contained in the {\sc iraf/stsdas}
package. This procedure is based on {the }five-level atom program
(de Robertis, Dufour \& Hunt 1987; Shaw \& Dufour 1995).

We compare these two sets of results {for} $t_3$, and get very
similar results, which are {shown} in Figure~\ref{fig.t3}.
 This means {that }these two methods from IRAF and {the
}literature are quite similar. In the following studies, we will
adopt the $t_3$ temperature estimated from {\sc temden.nebular}
in{ the} {\sc iraf/stsdas} package.

\begin{figure}
\centering
\includegraphics[angle=-90,width=8.8cm]{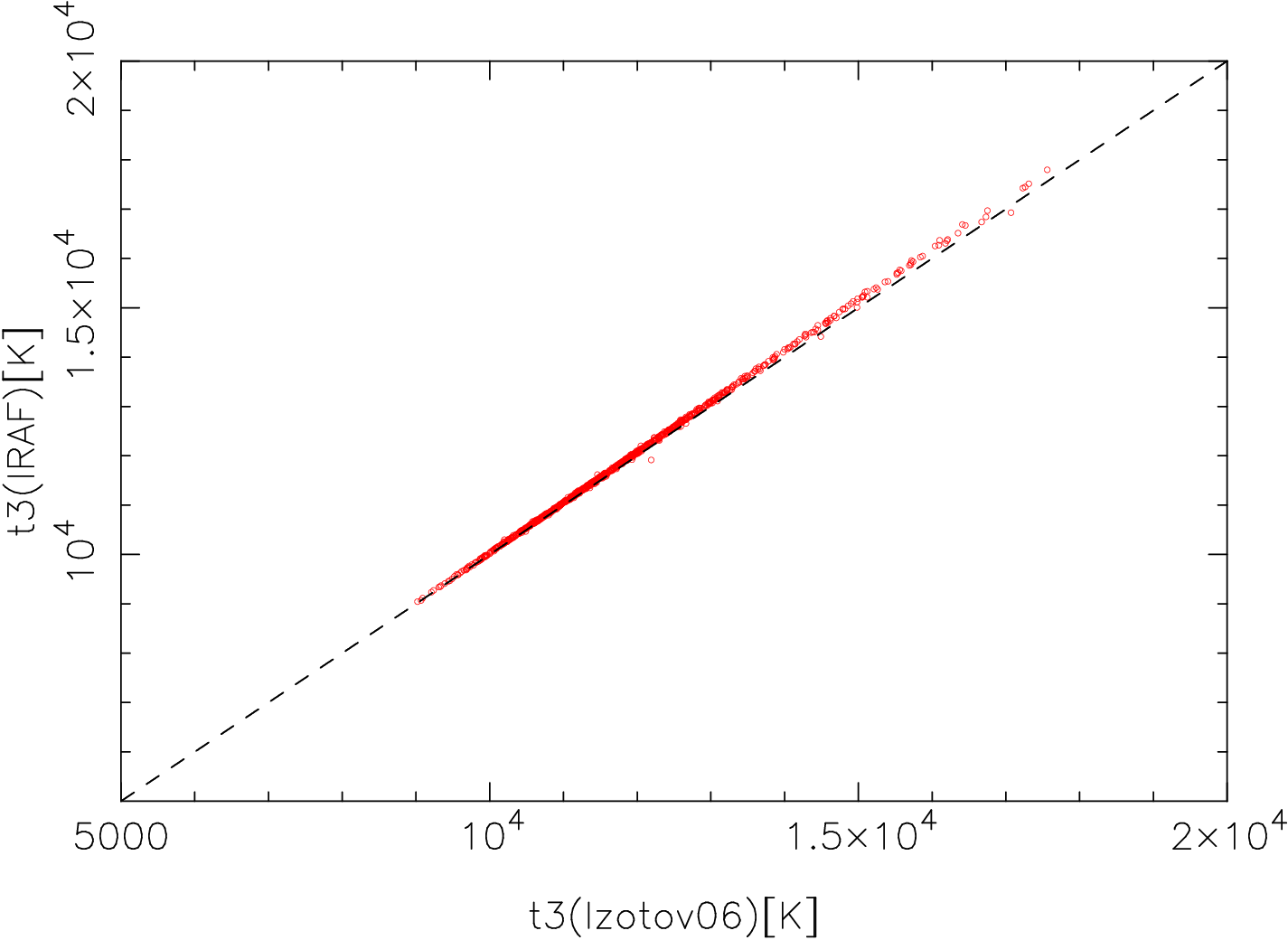}

\caption {\baselineskip 3.6mm Comparing the $t_3$ temperatures
estimated from the \cite{izo06} formula {with that from} IRAF ({\sc
temden.nebular}). They are very
  similar in most parts. Only in {the }high-temperature range
   is $t_3$(IRAF) 
    slightly
larger. The average offset is 71.8\,K from the equal-value line.}
\label{fig.t3}
\end{figure}

\subsubsection{Comparing $t_2$ from different analysis formulae}

To derive{ temperature} $t_2$, similar to \cite{gar92}, \cite{cam86}
used the photoionization models of \cite{sta82} to
derive a linear relation between $t_2$ and $t_3$
\begin{equation}
t_2 = 0.7t_3 + 0.3\,. \label{eq3}
\end{equation}
This is valid over the range 2000 ${\rm K}<T_{\rm e}$([O~{\sc
iii}])$<18\,000~ {\rm K}$, and has been widely used. \cite{pil10}
also derived a linear relation {between} $t_2-t_3$ as
\begin{equation}
t_2 = 0.835t_3 + 0.264\, . \label{eq4}
\end{equation}

They adopted the ff relation, and to convert the temperature
indicator into the electron temperature value $t_2$, they used the
five-level-atom solution of $\rm O^+$ with the Einstein coefficients
for spontaneous transitions A$_{jk}$ obtained by \cite{fro04} and
the effective cross{ }sections for electron impact $\Omega_{jk}$ from
\cite{pra06}.

\cite{izo06} also provide a set of analysis formulae to calculate
$t_2$ from $t_3$ in three different ranges of metallicities: ``low
$Z$" refers to $12+\log({\rm O/H})<7.2$, ``high $Z$" refers to
$12+\log({\rm O/H})>8.2$ and ``intermediate $Z$" refers to
$7.2<12+\log({\rm O/H})<8.2$.  Here we adopt their formulae for
intermediate $Z$ and high $Z$ to estimate $t_2${,} {but} {do not}
consider the  low $Z$ case.  These sequences were defined as in
\cite{sta03}, but have been recomputed with the latest atomic data
at that time, and with an input radiation field computed with
Starburst 99 (Leitherer et al. 1999) using the stellar model
atmospheres described in \cite{smi02}
\begin{eqnarray}
t_2&=-0.744+t_3\times(2.338-0.610t_3),& \quad {\rm intermed.\ }Z, \nonumber\\
   &=2.967+t_3\times(-4.797+2.827t_3),& \quad {\rm high\ }Z.
\label{toii}
\end{eqnarray}

These three transition formulae (Eqs.~(\ref{eq3}), (\ref{eq4}){
and} (\ref{toii})) can be used to estimate $t_2$ from $t_3$. We
would like to compare them by using the selected SDSS star-forming
galaxies in this work, which have detected [O~{\sc
iii}]$\lambda$4363 {that have been used to} estimate their
electron temperature, $t_3$. The results{ clearly} show the
difference {between} the three sets of{ temperature} $t_2$ in
Figure~\ref{fig.t2}, 
 where Line-1 refers to the result
from our Equation~(\ref{eq3}), Line-2 from Equation~(\ref{eq4}),
Line-3 from the intermediate-$Z$  case of Equation~(\ref{toii})
and Line-4 from its high-$Z$ case. Line-2 shows higher $t_2$ than
Line-1 at {a }given $t_3$. Line-3 (the intermediate-$Z$ case of
Eq.~(\ref{toii})) is close to Line-1 but{ it is} not a linear
relation.  Line-4 shows much discrepancy {compared to the} other
three lines except{ at} the low temperature end. Thus we would not
take Line-4 to estimate abundances in the later part of this work.
We will estimate $\rm O^{+}$/$\rm H^+$ abundances of our sample
galaxies by using{ the temperature} $t_2$ shown by Line{s}-1,{ }2{
and }3 in {the }next section.

We keep in mind that, as mentioned in \cite{and13}, the $t_2-t_3$
relation {from} our Equation~(\ref{eq3}) overestimates the low
ionization zone $T_{\rm e}$[O~{\sc ii}] by $\sim$1000--2000\,K and
will underestimate the oxygen abundance by $\sim$0.18\,dex. In
this work, we {can}not carefully check this result since {by using
}the {spectra} of these individual objects}{, we} cannot reliably
estimate fluxes of their [O~{\sc ii}]$\lambda\lambda$7320,7330
auroral lines. {However,} in Section~3.4 we will discuss such
effects on discrepancies in temperature and abundances following{
the} suggestions of \cite{and13}.

\begin{figure}
\centering
\includegraphics[angle=-90,width=8.8cm]{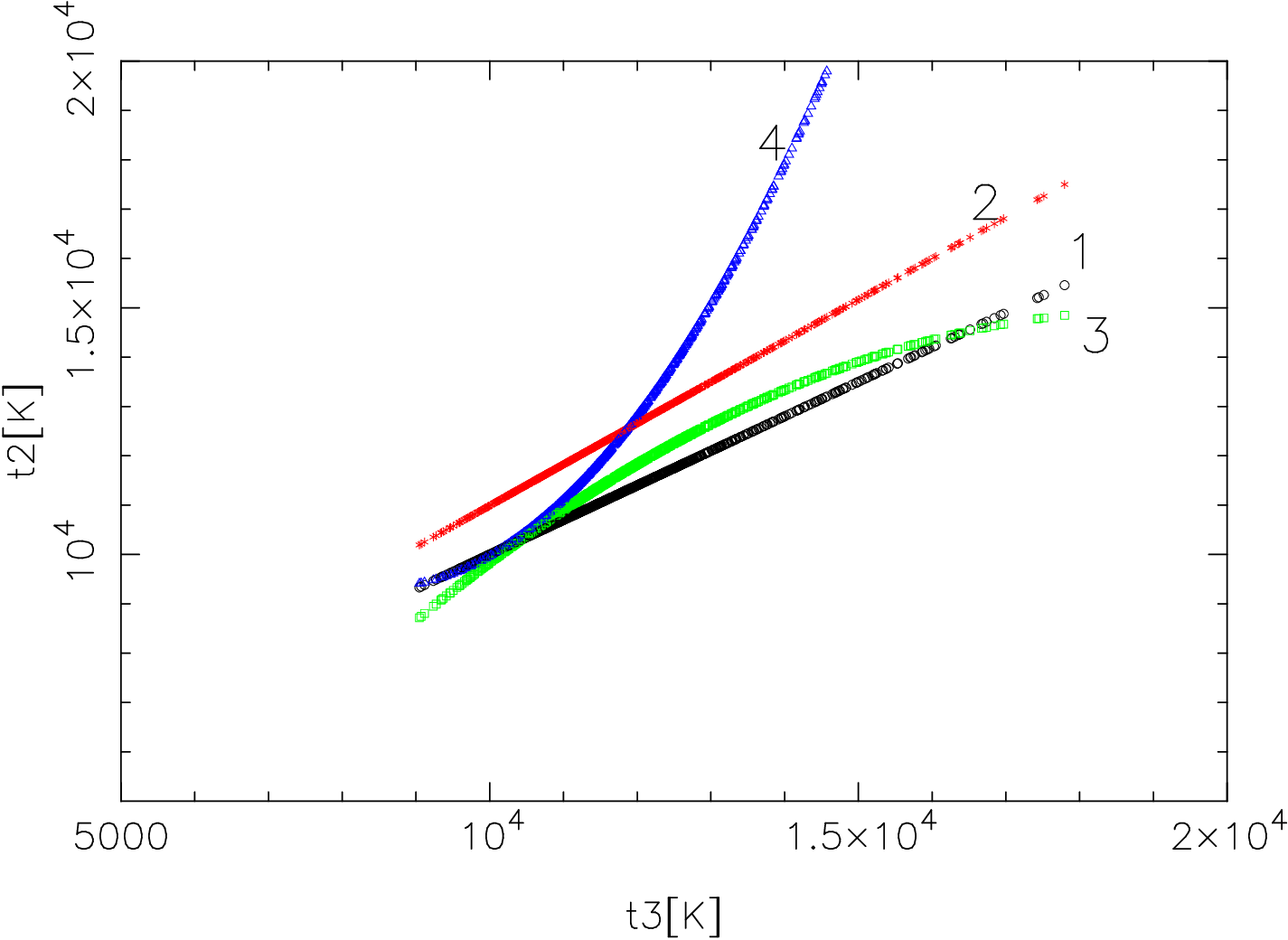}

\caption {\baselineskip 3.6mm Comparison of the $t_2$ temperature
estimated from $t_3$ using
  the three transition formulae: Line-1 refers to the result from
  Eq.~(\ref{eq3}), Line-2 from Eq.~(\ref{eq4}), Line-3 from the
  intermediate-$Z$ case of Eq.~(\ref{toii}) and Line-4 from its
  high-$Z$  case. The data points are the star-forming galaxies from SDSS
  {whose $t_3$ values}   have been estimated from [O~{\sc
    iii}]$\lambda\lambda$4959, 5007/[O~{\sc iii}]$\lambda$4363 with IRAF.}
\label{fig.t2}
\end{figure}

\subsection{Comparing the Ionic Abundances from Two Different Methods}

The ionic oxygen abundances $\rm O^{+}$/$\rm H^+$ and $\rm
O^{++}$/$\rm H^+$ can be estimated from the electron temperature
$t_2${ and} $t_3$ and the related emission line
 ratios, respectively. The
total oxygen abundances in the ionized nebulae 
 can be
derived from the relation of the line ratios
\begin{equation}
\frac{\rm O}{\rm H} = \frac{\rm O^+}{\rm H^+} + \frac{\rm
O^{++}}{\rm H^+}\,.
\end{equation}

The contributions from other ions such as O$^{3+}$ can be omitted
since they are small.
 \cite{izo06} have recently published a set of
equations for the determination of the oxygen abundances in H {\sc
ii} regions for a five-level atom. They used the atomic data from
the references listed in \cite{sta05}. They show the ionic
abundances $\rm O^{++}$/$\rm H^+$ and $\rm O^{+}$/$\rm H^+$ are
estimated as follows:
\begin{eqnarray}
12 + \log (\rm O^{++}/\rm H^+) &=&
\log(I_{\rm {[O}{ III]\lambda4959+\lambda5007}}/I_{\rm H\beta})  \nonumber \\
               &&+ 6.200 + \frac{1.251}{t_3} - 0.55\log t_3 - 0.014t_3 \, ,
\\
12 + \log(\rm O^{+}/\rm H^+) &=&
\log(I_{\rm {[O}{\rm II]\lambda3726+\lambda3729}}/I_{\rm H\beta}) +  5.961   \nonumber \\
        &&+ \frac{1.676}{t_2} - 0.40\log t_2 - 0.034
        t_2 + \log (1 + 1.35x_2)\, ,
\end{eqnarray}
with $x_2 = 10^{-4}n_{\rm e}t_2^{-1/2}$, and {$n_{\rm
e}$} is the electron density in cm$^{-3}$.

The ionic abundances $\rm O^{++}$/$\rm H^+$ and $\rm O^+$/$\rm H^+$
can also be estimated by using the task {\sc ionic.nebular}
contained in the {\sc
  iraf/stsdas} package. We compare the ionic abundances from these
two methods.

\begin{figure}
\vs
\begin{minipage}{0.5\textwidth}
\centering
\includegraphics[angle=-90,width=70mm]{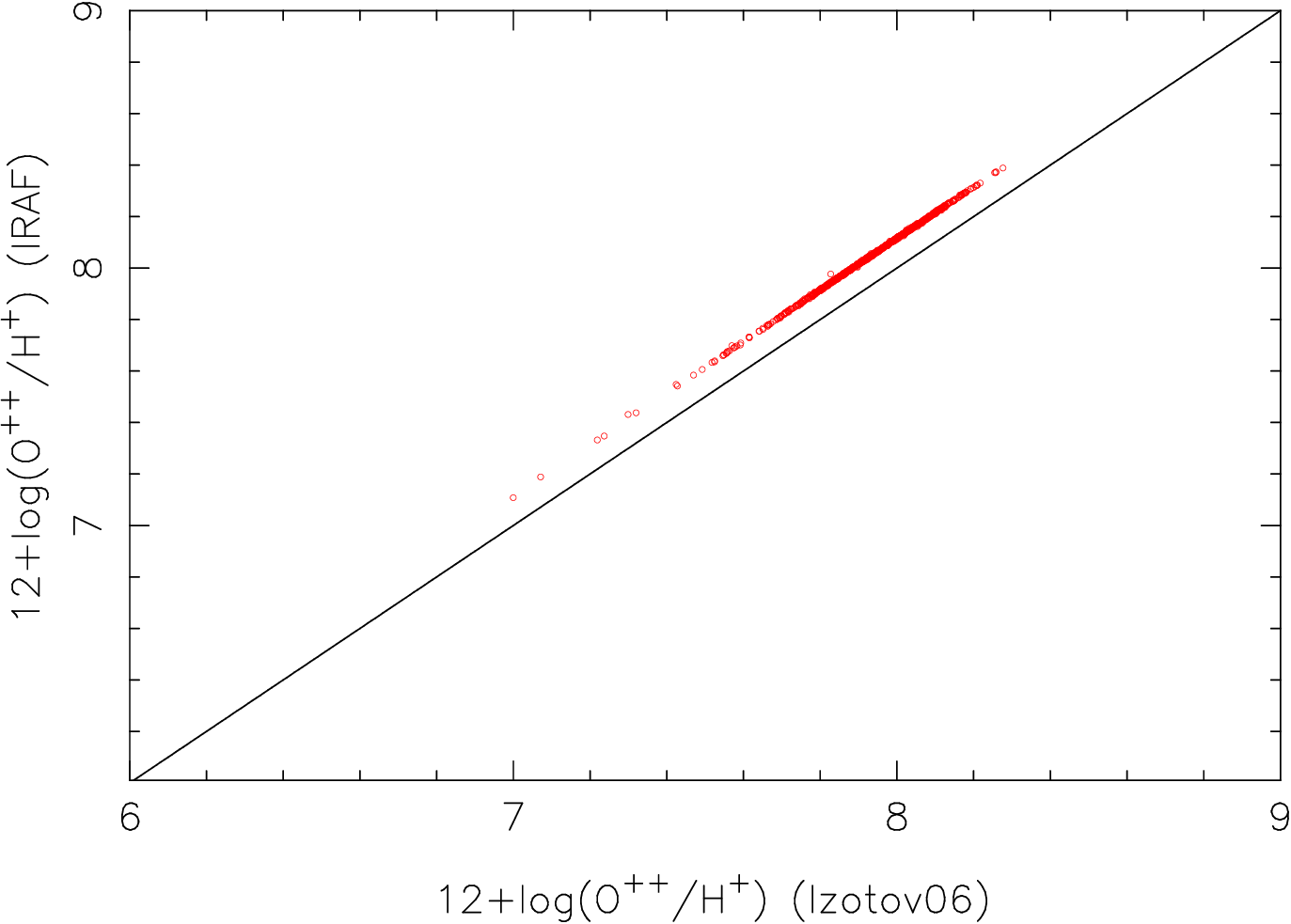}
\end{minipage}
\begin{minipage}{0.5\textwidth}
\centering
\includegraphics[angle=-90,width=70mm]{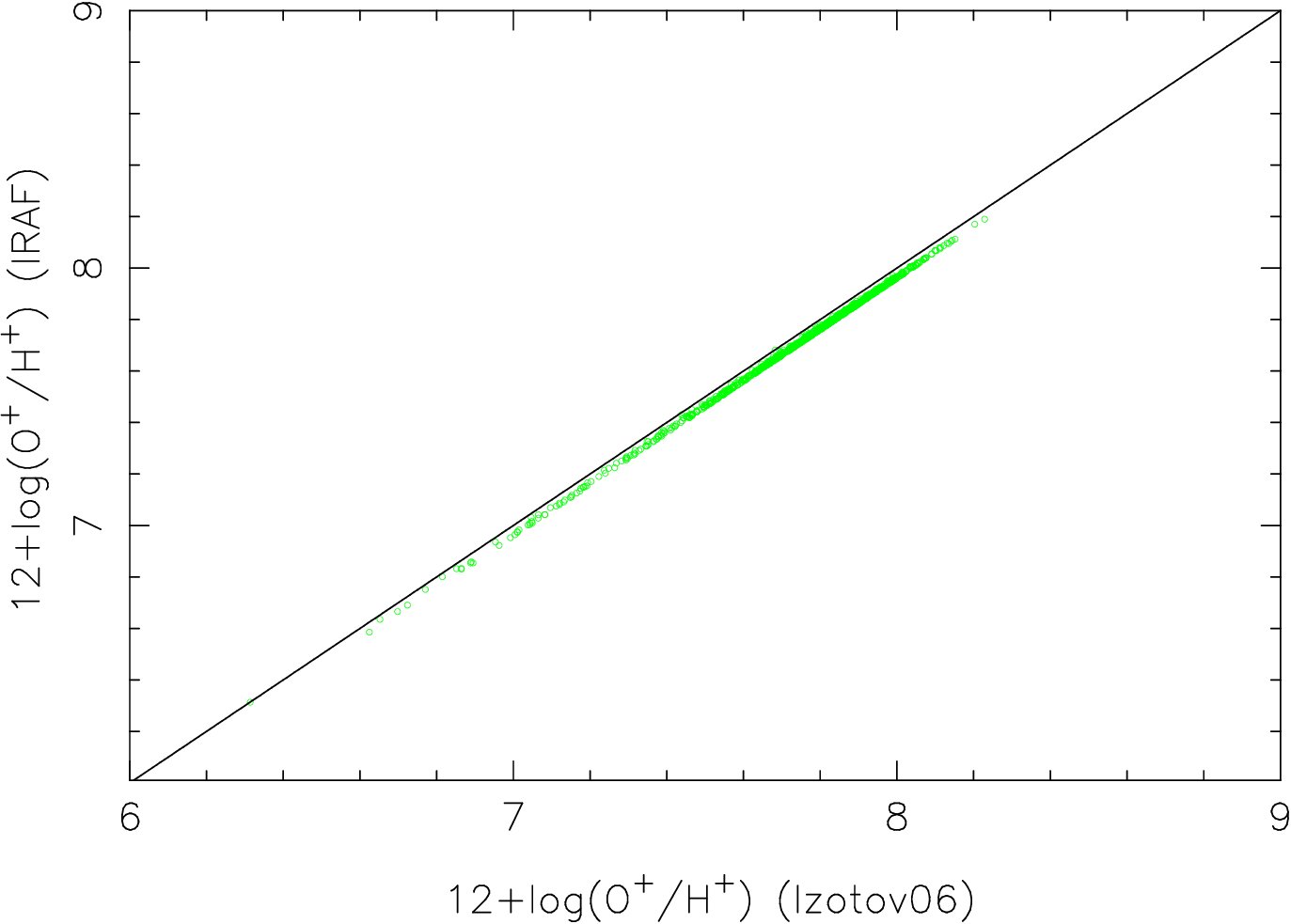}
\end{minipage}
\caption {\baselineskip 3.6mm {\it Left}: Compari{so}n{ of} the $\rm
O^{++}$/$\rm H^+$ abundances from
  {\sc ionic.nebular} and \cite{izo06} ($t_3$ from IRAF). We can see the abundances from
  {\sc ionic.nebular} are {slightly} bigger. The average deviation is
  0.11\,dex from the equal-value line. {\it Right}: Compari{so}n{ of} the $\rm O^{+}$/$\rm H^+$ abundances from
  {{\sc ionic.nebular}} and \cite{izo06} ($t_2$ from Equation $t_2=0.7t_3+0.3$). The abundances
  are very similar for the two methods. The average deviation is
  0.03\,dex from the equal-value line. }
\label{fig.o3.o2}
\end{figure}

The $\rm O^{++}$/$\rm H^+$ abundances from {\sc ionic.nebular} and
\cite{izo06} are compared in the left panel of
Figure~\ref{fig.o3.o2} 
 ($t_3$ is calculated from {\sc
temden.nebular}). IRAF gives {a }slightly higher abundance, about
0.1\,dex. The right panel of Figure~\ref{fig.o3.o2} is for $\rm
O^{+}$/$\rm H^+$ abundances{;} here we adopt $t_2=0.7t_3+0.3$ to
estimate $t_2$, and then the derived $\rm O^{+}$/$\rm H^+$
abundances from {\sc ionic.nebular} and \cite{izo06} are also
compared. These results show that the ionic abundances from{ the}
\cite{izo06} formula are quite similar to those from{ the} 
IRAF/IONIC procedure,{ and} the average offset is 0.03\,dex. The
electronic temperatures and oxygen abundances of 947 sample
galaxies are given in Table~\ref{tab2}.

\subsection{Comparing the $\rm O^+$/$\rm H^+$ Abundances from Different $t_2$ }

Now we compare the $\rm O^{+}$/$\rm H^+$ abundances from the three
different sets of $t_2$ from \cite{gar92} (Eq.~(\ref{eq3})),
\cite{pil10} (Eq.~(\ref{eq4})) and \cite{izo06} (Eq.~(\ref{toii})).

\begin{figure*}
\begin{minipage}{0.5\textwidth}
\centering
\includegraphics[angle=-90,width=70mm]{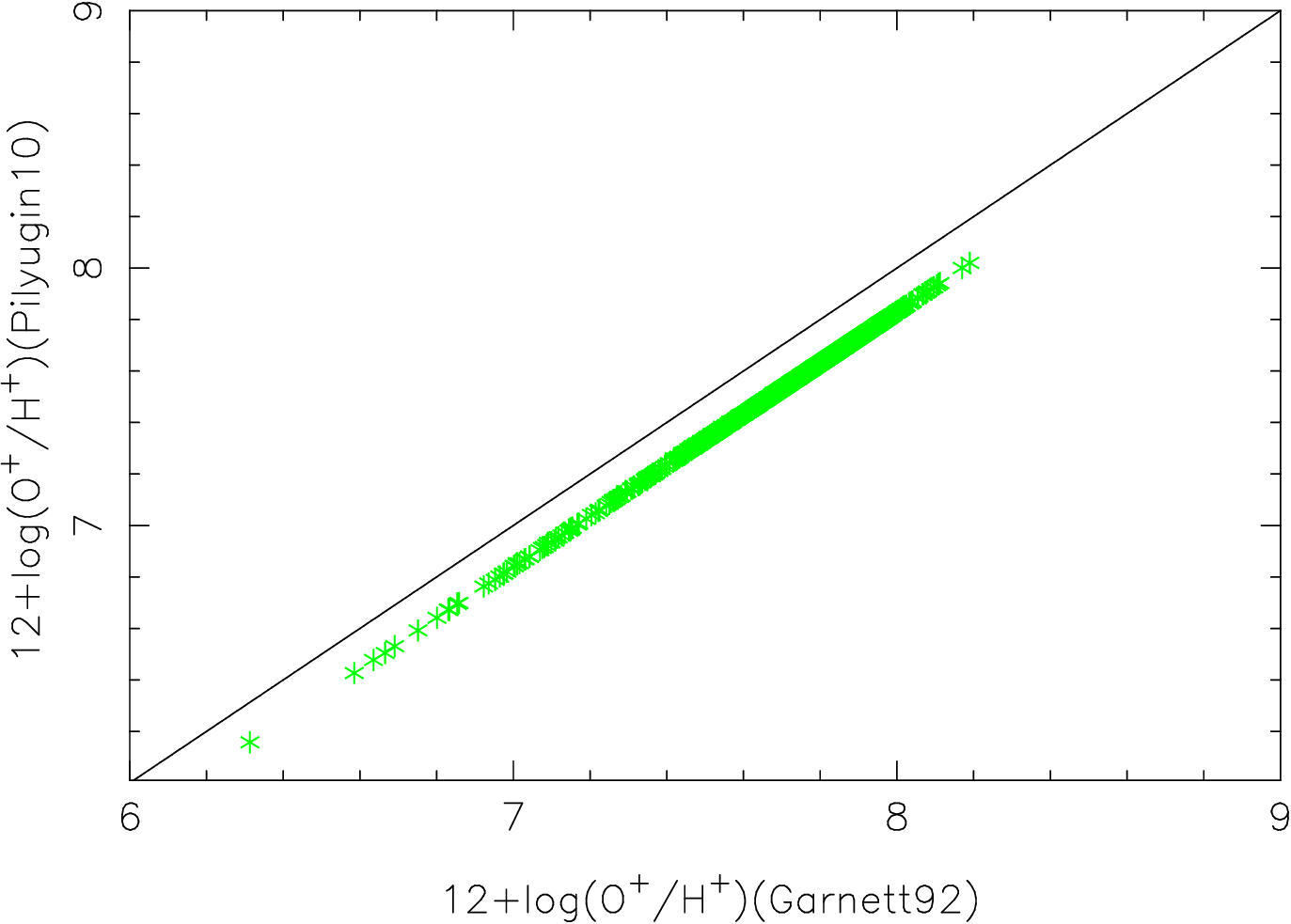}
\end{minipage}
\begin{minipage}{0.5\textwidth}
\centering
\includegraphics[angle=-90,width=70mm]{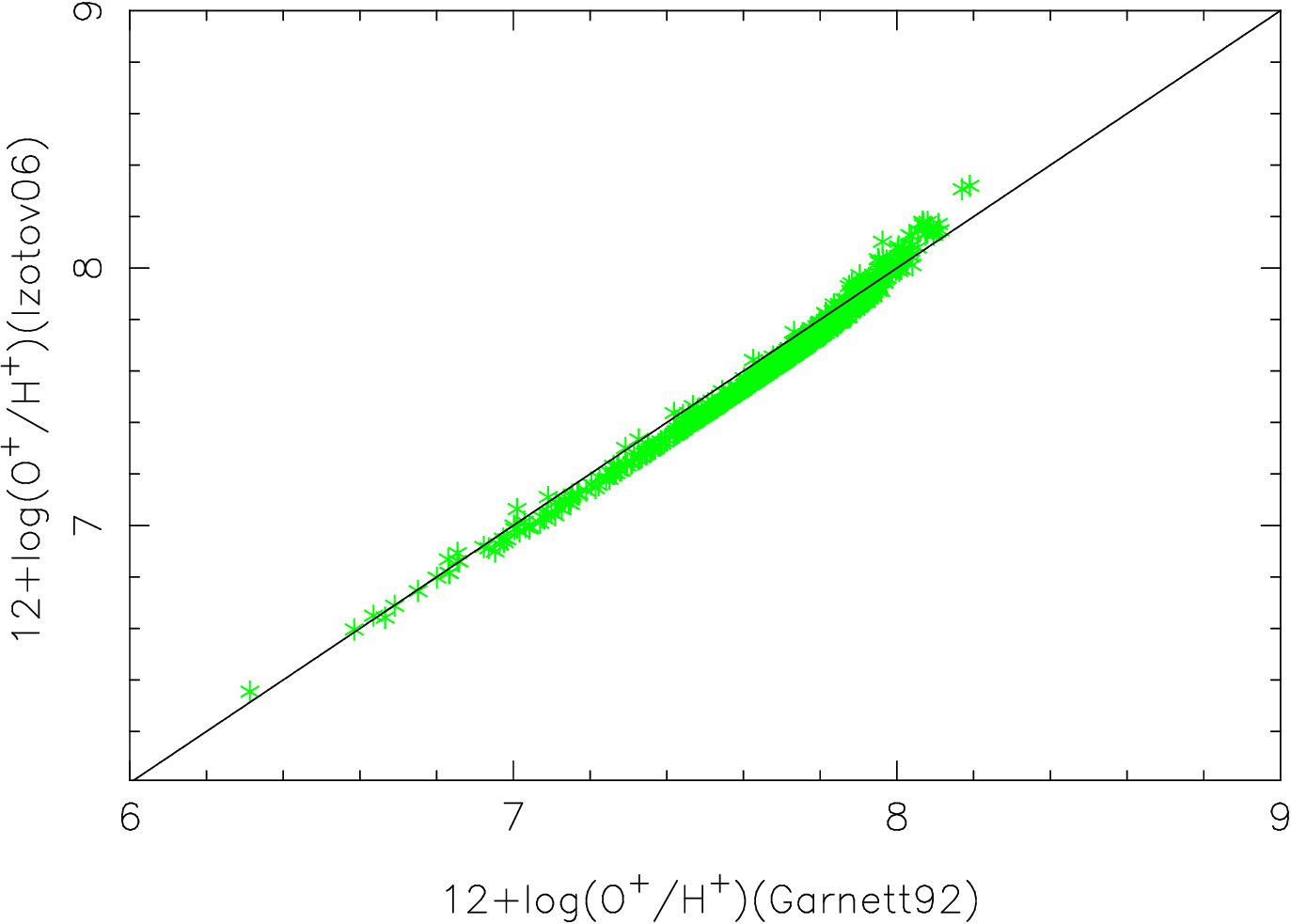}
\end{minipage}
\caption {\baselineskip 3.6mm {\it Left}: Compari{so}n{ of} the $\rm
O^{+}$/$\rm H^+$ abundances with different $t_2$
  formulae from \cite{gar92} (Eq.~(\ref{eq3})) and \cite{pil10} (Eq.~(\ref{eq4})). The average residual is
  0.15\,dex. {\it Right}: Compari{so}n{ of} the $\rm O^{+}$/$\rm H^+$ abundances with different
  $t_2$ formulae from \cite{gar92} (Eq.~(\ref{eq3})) and \cite{izo06} (Eq.~(\ref{toii}), intermediate-Z
  case). The average residual is 0.05\,dex. The derived $\rm O^{+}$/$\rm H^+$
  abundances are from {\sc ionic.nebular}.}
\label{fig.o22}
\end{figure*}

The derived $\rm O^{+}$/$\rm H^+$ abundances are derived using
{the }{\sc ionic.nebular} procedure. The results are compared in
Figure~\ref{fig.o22}. 
  The $\rm O^{+}$/$\rm H^+$
abundances from \cite{izo06} are quite similar to the abundances
from \cite{gar92}, and the one from \cite{pil10} will result in
about 0.15\,dex lower $12+\log\rm (O^+/H^+)$ abundances than that
of \cite{gar92}.

\subsection{Comparing the $\rm O^{++}$/$\rm H^+$ and the $\rm O^+$/$\rm H^+$ Abundances }

\begin{figure}

\vs\vs
\begin{minipage}{0.5\textwidth}
\centering
\includegraphics[angle=-90,width=70mm]{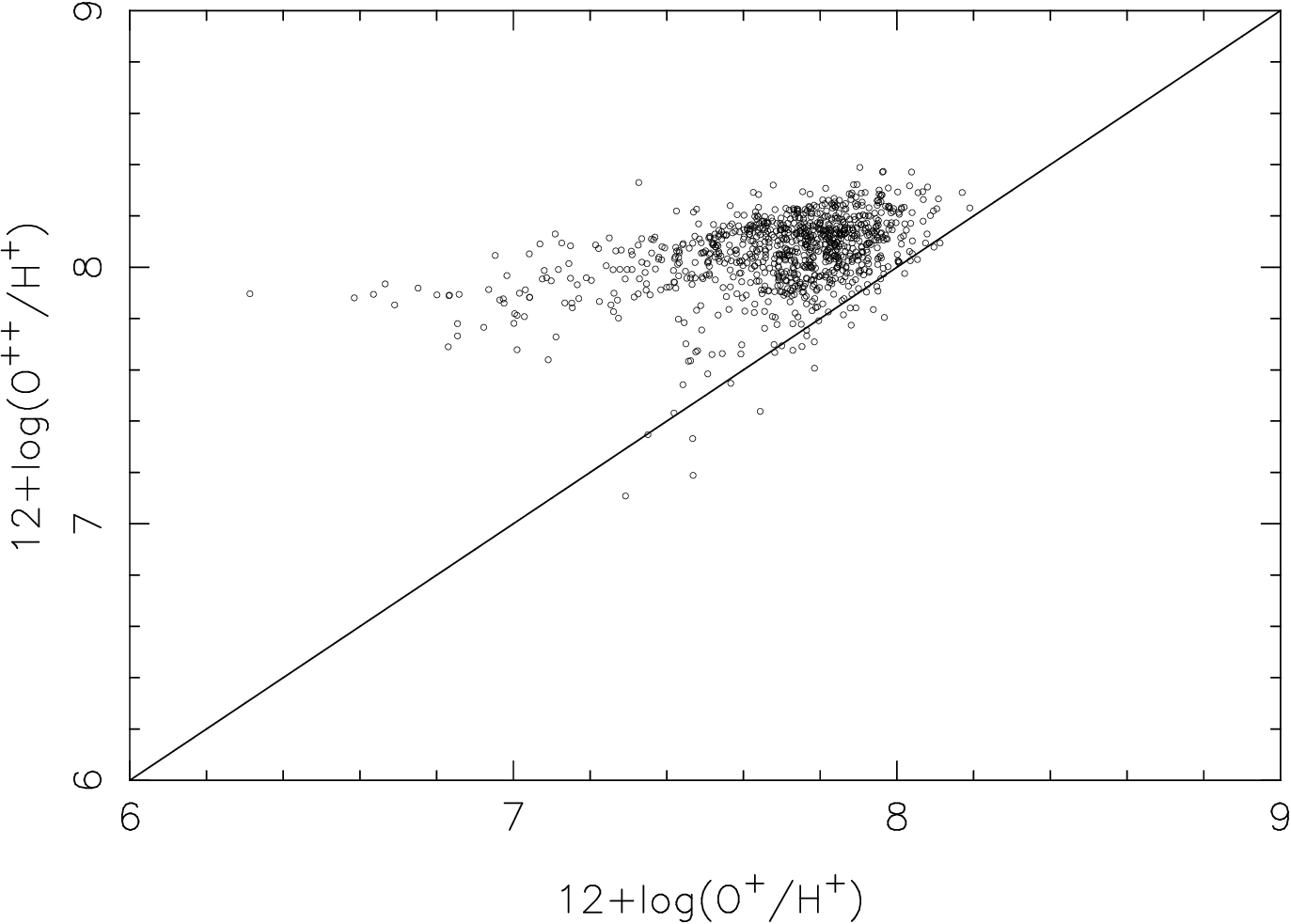}
\end{minipage}
\begin{minipage}{0.5\textwidth}
\centering
\includegraphics[angle=-90,width=68mm]{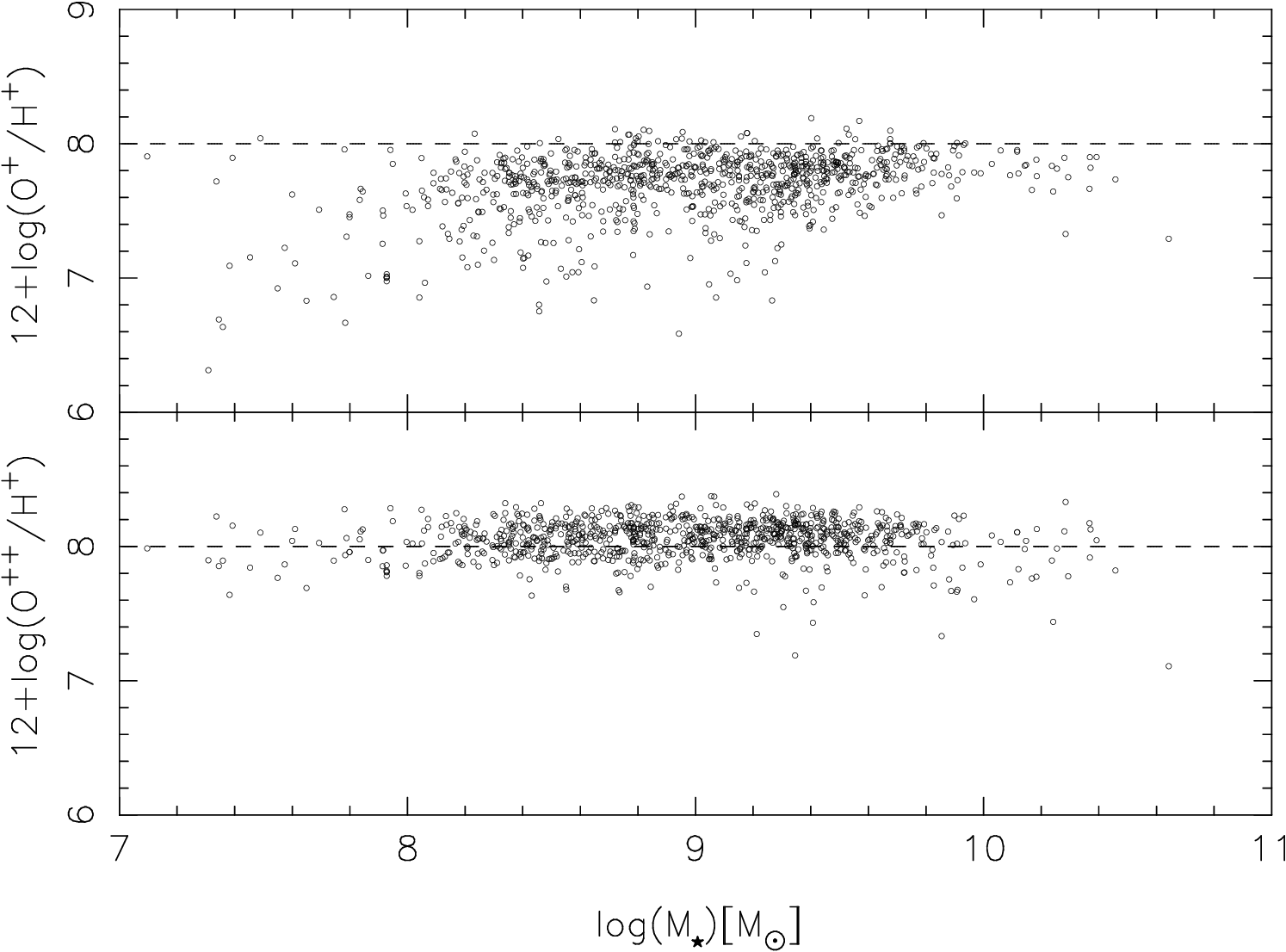}
\end{minipage}

\vs
\begin{minipage}{0.5\textwidth}
\centering
\includegraphics[angle=-90,width=70mm]{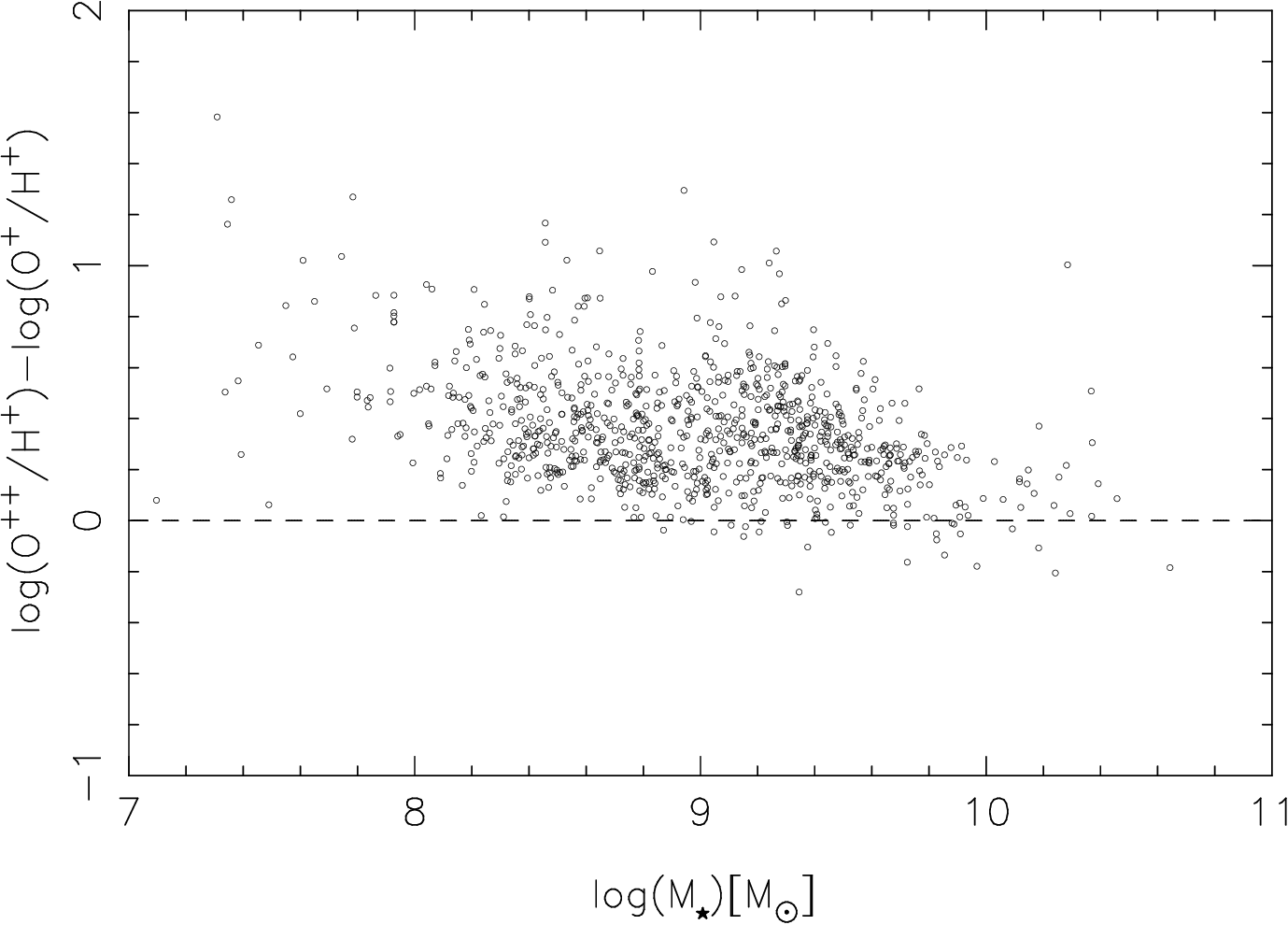}
\end{minipage}
\begin{minipage}{0.5\textwidth}
\centering
\includegraphics[angle=-90,width=70mm]{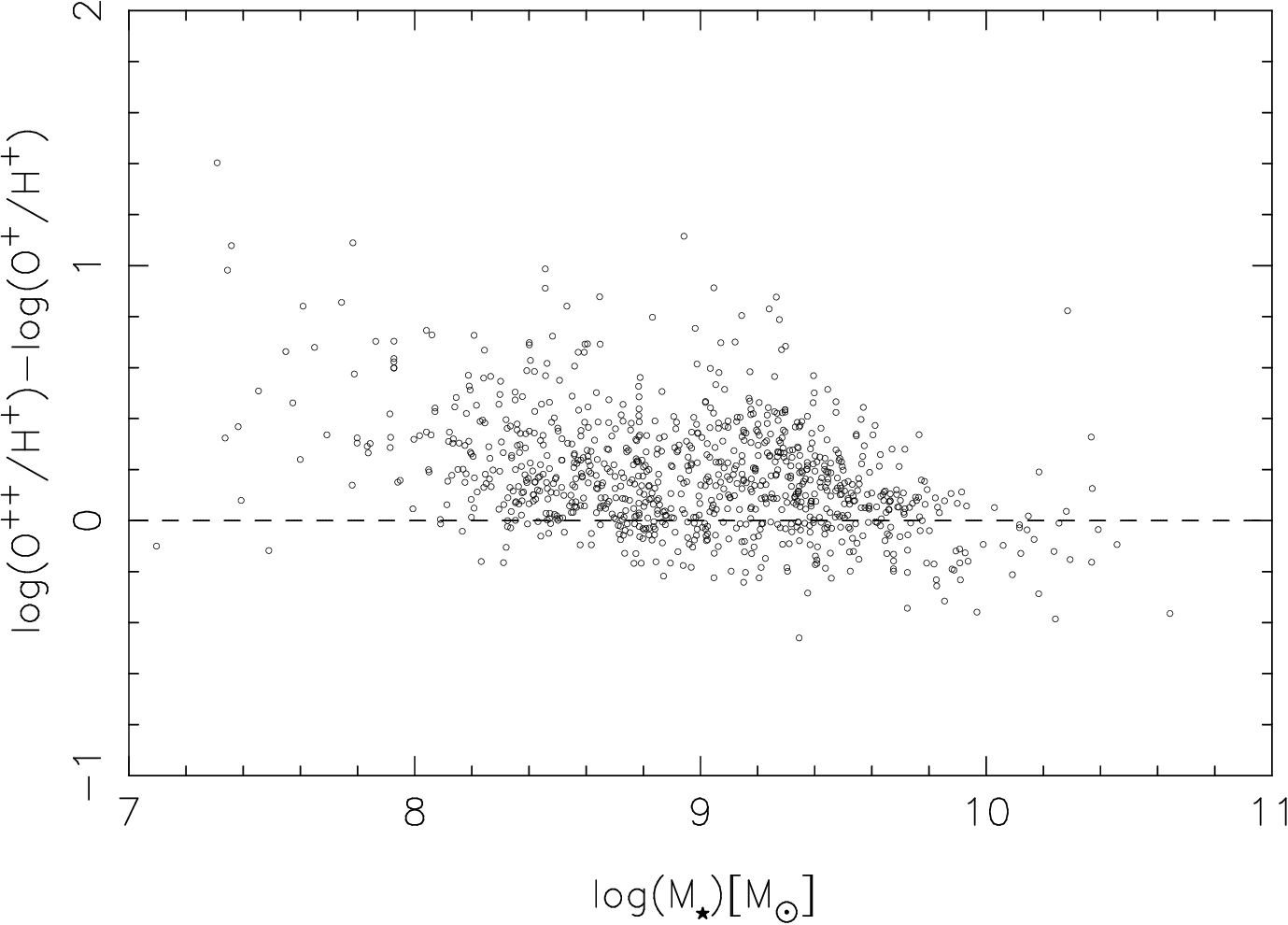}
\end{minipage}

\caption {\baselineskip 3.6mm {\it {Upper left}}:
Comparing the $\rm O^{++}$/$\rm H^+$ and $\rm O^{+}$/$\rm H^+$
  abundances. {\it {Upper right}}: Showing the relations $\rm O^{++}$/$\rm H^+$
  vs. stellar mass and $\rm O^{+}$/$\rm H^+$ vs. stellar mass.  For
  comparison, we also plot the equal-value (8.0) line with a dashed
  line.  {\it {Lower left}}: The relations of stellar mass vs. the residuals between $\rm O^{++}$/$\rm H^+$ and
  $\rm O^+$/$\rm H^+$. The average offset of the residuals is 0.36\,dex from the
  zero-point line.  {\it {Lower right}}: Similar relations of stellar mass vs. the residuals between $\rm O^{++}$/$\rm H^+$
  and $\rm O^+$/$\rm H^+$, but the $\rm O^+$/$\rm H^+$ abundances have been artificially
  increased by 0.18\,dex according to suggestions of \cite{and13}. These plots show that
  O$^{++}/\rm H^+$ is larger than $\rm O^+$/$\rm H^+$ for most galaxies. The{ values for} electron temperature $t_2$
  are estimated using the formula from
  \cite{gar92}{ }(Eq.~(\ref{eq3})){.} $\rm O^{++}$/$\rm H^+$ and $\rm O^+$/$\rm H^+$ oxygen abundances are calculated
  from {\sc ionic.nebular}.}
\label{fig.oo}
\end{figure}

Now we compare the ionic abundances $\rm O^{++}$/$\rm H^+$ and
$\rm O^{+}$/$\rm H^+$ in Figure~\ref{fig.oo} 
 in the
relation with stellar mass{;} here we adopt the $\rm O^{++}$/$\rm
H^+$ and
  $\rm O^{+}$/$\rm H^+$ from {\sc ionic.nebular} in {\sc IRAF} and $t_2=0.7t_3+0.3$ for{ temperature} $t_2$.

Figure~\ref{fig.oo} shows that for most cases $\rm O^{++}$/$\rm H^+$
is higher than $\rm O^{+}$/$\rm H^+$. We did not see the suggestions
by \cite{and13} that $\rm O^{+}$/$\rm H^+$ is more significant than
$\rm O^{++}$/$\rm H^+$ (their fig.~5). Maybe that is because they
are focusing on the massive galaxies. Another reason could be that
we may underestimate the $\rm O^{+}/\rm H^+$ abundances since the
analysis formula we used {is} $t_2=0.7t_3+0.3${ which} overestimates $t_2$ as
mentioned in \cite{and13} (their fig.~5). This effect is shown in
the {lower right} panel of
Figure~\ref{fig.oo}.

\subsection{Comparing the Oxygen Abundances from {the }$T_{\rm e}$ Method and $R_{23}$ Method}

\begin{figure*}
\begin{minipage}{0.5\textwidth}
\centering
\includegraphics[angle=-90,width=70mm]{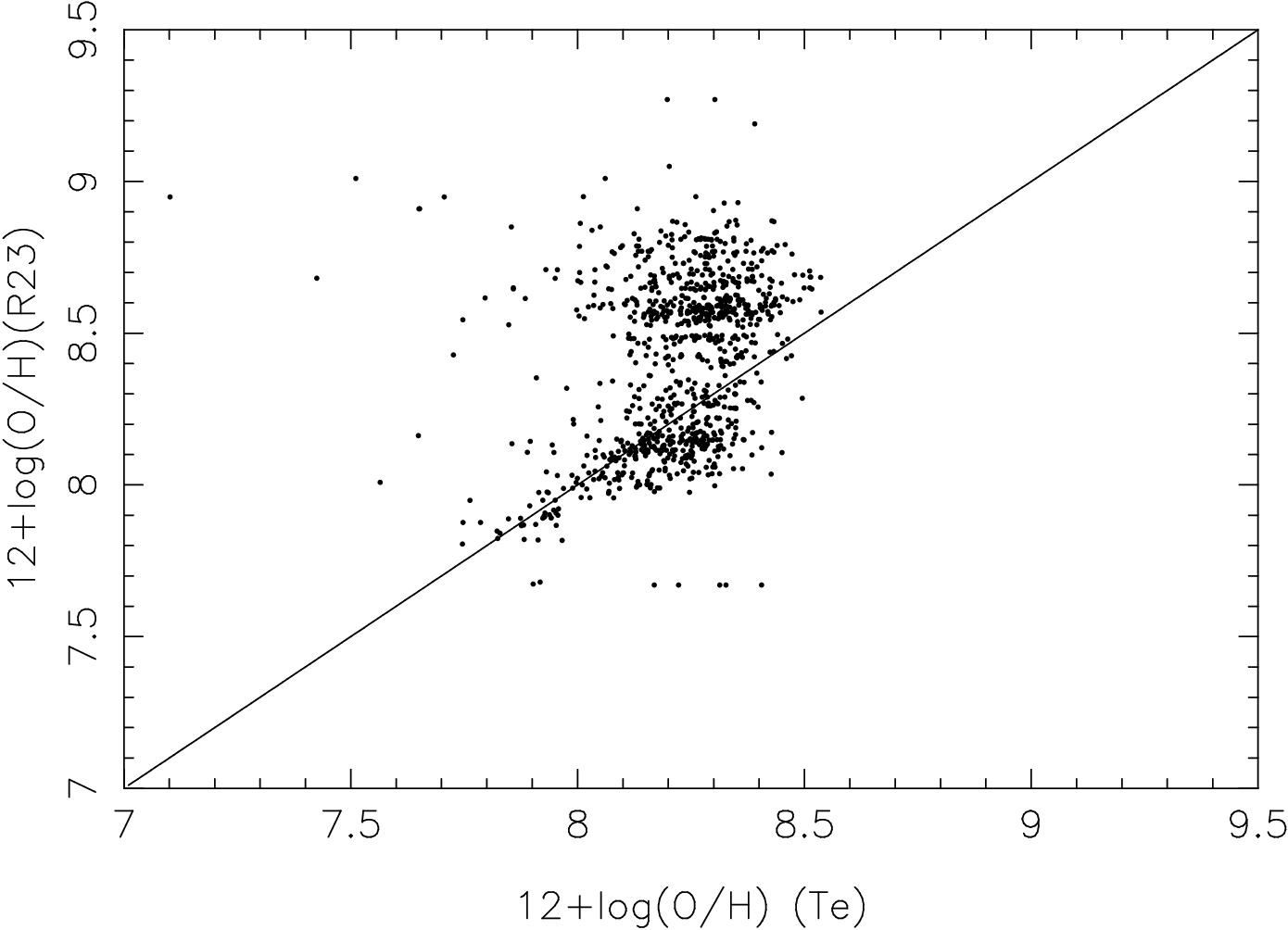}
\end{minipage}
\begin{minipage}{0.5\textwidth}
\centering
\includegraphics[angle=-90,width=70mm]{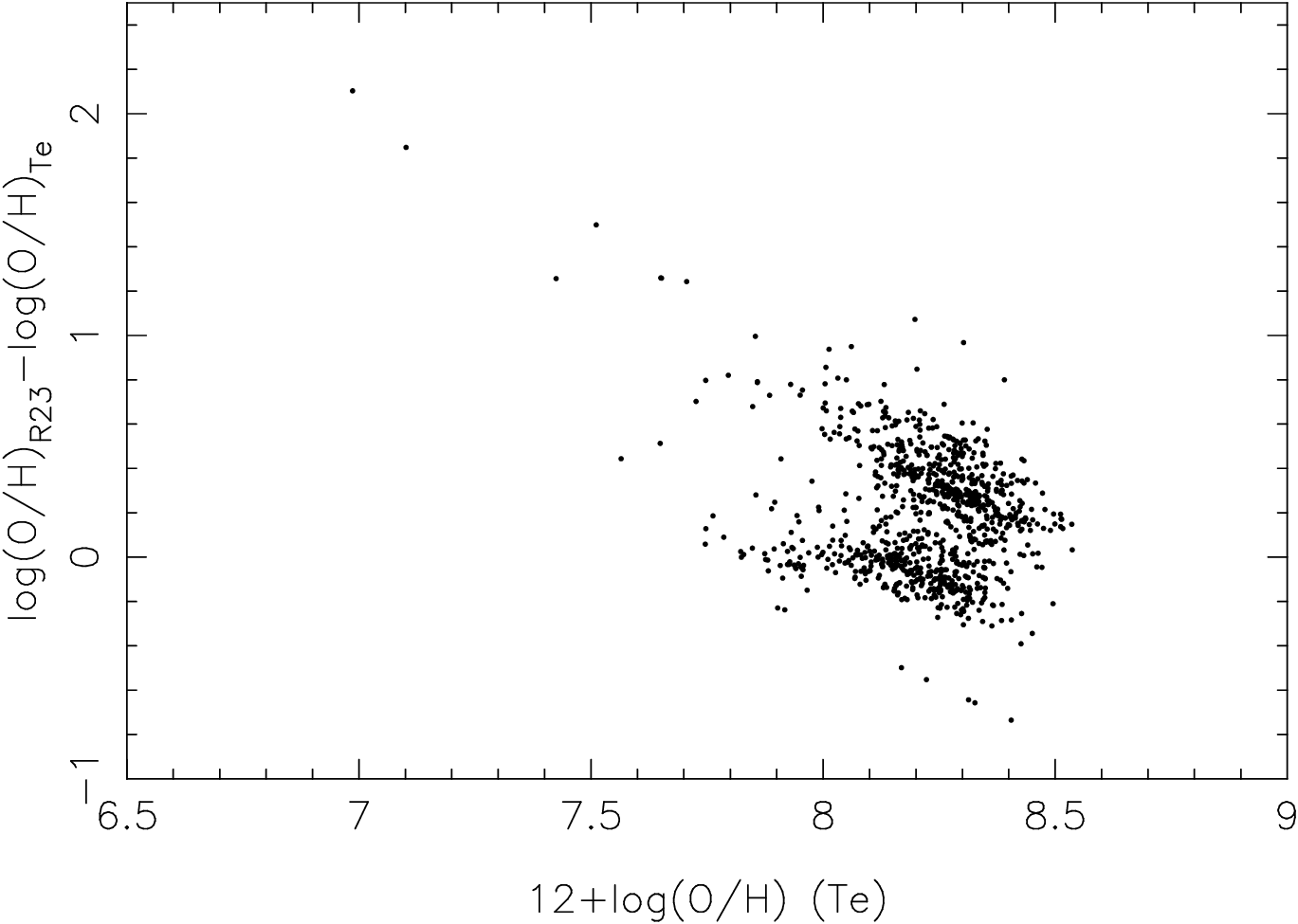}
\end{minipage}
\begin{minipage}{0.5\textwidth}
\centering
\includegraphics[angle=-90,width=70mm]{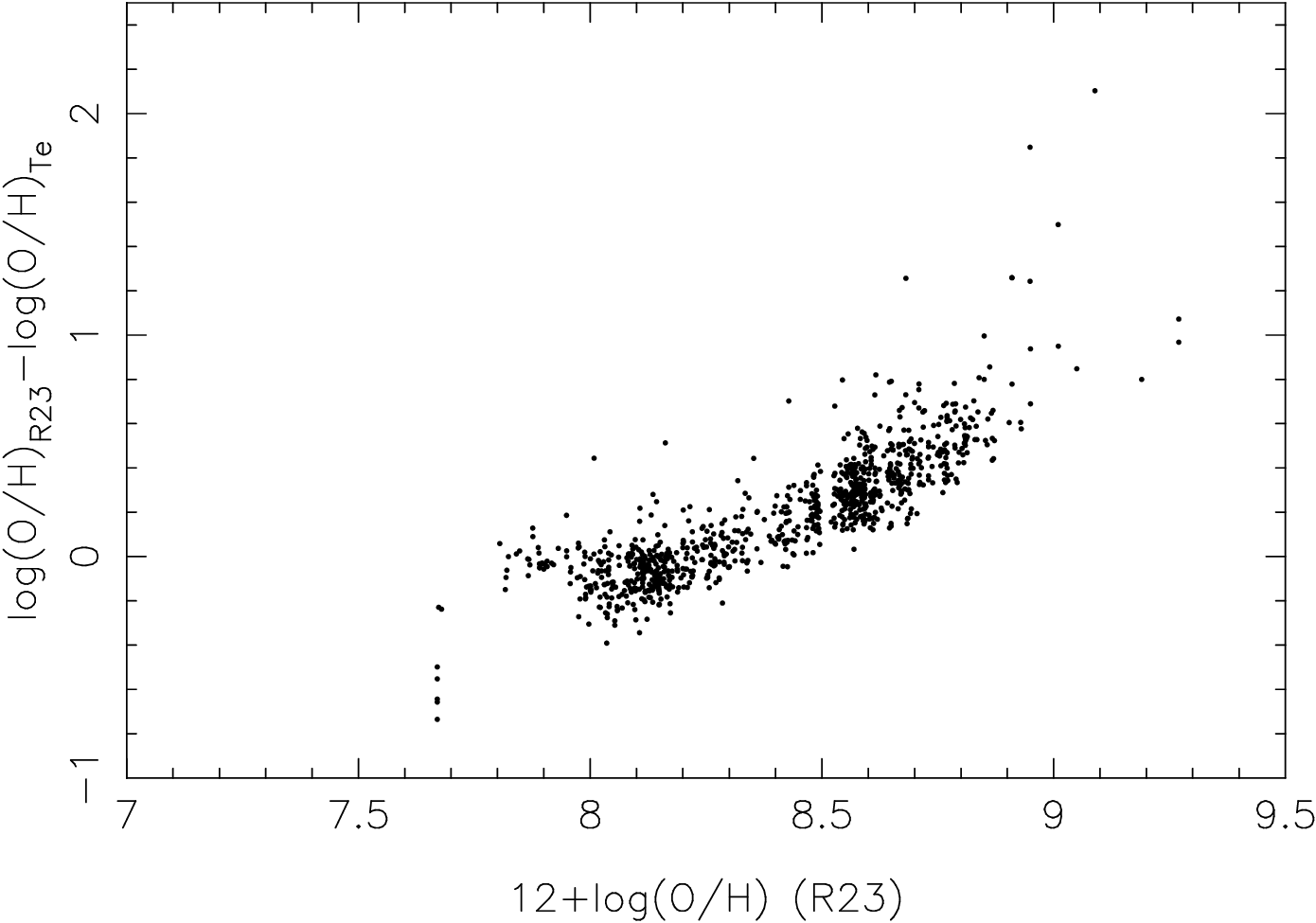}
\end{minipage}
\begin{minipage}{0.5\textwidth}
\centering
\includegraphics[angle=-90,width=70mm]{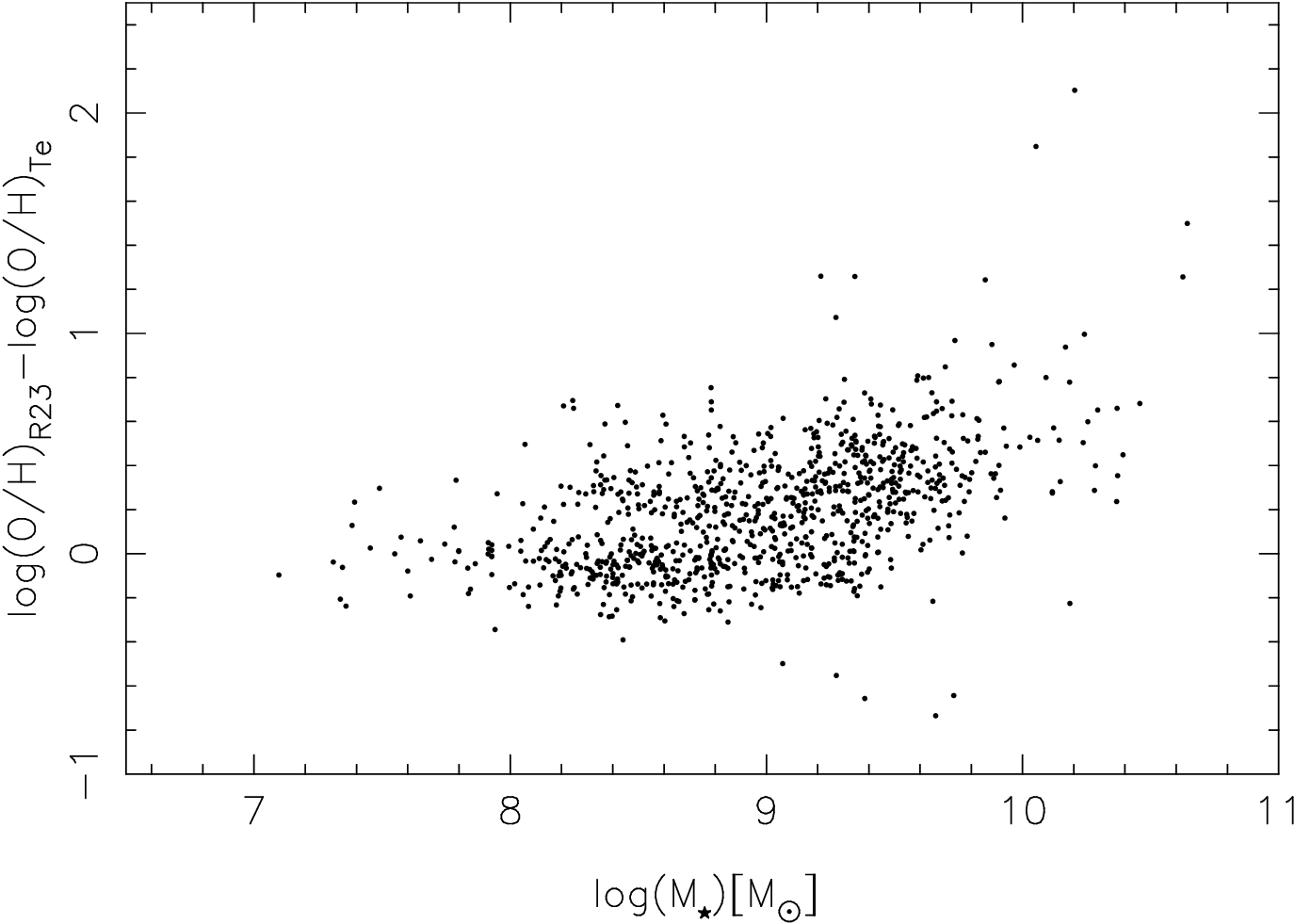}
\end{minipage}
\caption{\baselineskip 3.6mm Comparing the oxygen abundances derived
from the $T_{\rm e}$ method
  and from the $R_{23}$ method. The residuals from the two methods
  vs. $12+\log({\rm O/H})(T_{\rm e}$), vs. $12+\log ({\rm O/H})(R_{23})$ and vs. $\log
  (M_*)$
  are also shown. The sample galaxies are divided {in}to two populations
  and the residuals become bigger with {an} increase {in}
  $12+\log({\rm O/H}) (R_{23}$) and $\log (M_*)$. }
\label{fig.oh.com11}
\end{figure*}

We compare the difference between the oxygen abundances
$12+\log$(O/H) obtained from{ the} electronic temperature method
($T_{\rm e}$-method) and the strong emission-line ratio method
(such $R_{23}$). This is {shown} in Figure~\ref{fig.oh.com11}.

We adopt the oxygen abundance based on the strong-line method
obtained by {the }MPA/JHU group. The MPA/JHU group used{ a}
photoionization model to simultaneously fit most prominent
emission lines. Based on {a} Bayesian technique, they have
calculated the likelihood distribution of the metallicity of each
galaxy in the sample by comparing with a large library of models
corresponding to different assumptions about the effective gas
parameters, and then they adopted the median of the distribution
as the best estimate of the metallicity{ of the galaxy} (Yin et
al. 2007; Tremonti et al. 2004; Brinchmann et al. 2004). These are
quite similar to the abundances from {the }$R_{23}$ method, e.g.,
the strong line ratio of ([O~{\sc ii}]3727+[O~{\sc iii}]4959,
5007)/$\rm H\beta$.

The differences {between} {the }$T_{\rm e}$ method and $R_{23}$
method are quite clear{,} as shown in Figure~\ref{fig.oh.com11}.
 For {a }significant part of the galaxies, the $R_{23}$
method give{s} much higher oxygen abundances than the $T_{\rm e}$
method. It seems that the $T_{\rm e}$-based oxygen
  abundances have a{n} upper limit{ of} $12+\log {\rm (O/H)}$ around 8.5. Above this
  metallicity, it is very difficult to excite the collision excitation
  line [O~{\sc iii}]$\lambda$4363 in the ionized nebulae, thus it
  becomes very difficult to obtain their electron temperature and the
  direct oxygen abundances. More physical scenario{s} can be found in
  Section~4.1.  {I}t{ also} seems the sample galaxies are divided into two
populations following $12+\log {\rm (O/H)}_{R_{23}}$. \cite{yin07}
ha{ve} explained the reason in their {s}ection~4.1, which could
possibly{ be} related to how secondary nitrogen enrichment is
treated in the \cite{cl01} models which MPA/JHU used to derive the
metallicities of the galaxies. The discrepancy between the oxygen
abundances from {the }$R_{23}$ and $T_{\rm e}$ methods {is} more
obvious {with} increasing stellar masses {in} the galaxies.

The differences between abundances from {the }$R_{23}$ and $T_{\rm
e}$ methods are larger following the increasing stellar masses
although the scatters are obvious. This could be due to the bias
of the sample themselves, i.e. lower metallicity ones were
selected relative to the cases of more massive galaxies{;} Or
perhaps this discrepancy {is} consistent with the suggestion
in the abstract of Stasinska (2005): We find that, for
metallicities larger than solar, the computed abundances deviate
systematically from the real ones, generally by larger amounts for
more metal-rich 
 H~{\sc ii} regions.

\section{Deriving the mass-metallicity relations of galaxies from
{$T_{\lowercase {\rm e}}$}-based oxygen abundances}

\subsection{Can the [O~{\sc iii}]$\lambda$4363
 Selected Galaxies be Good Representatives of the Whole{ Population of} Galax{ies?}}

\begin{figure*}

\vs
\begin{minipage}{0.5\textwidth}
\centering
\includegraphics[angle=-90,width=70mm]{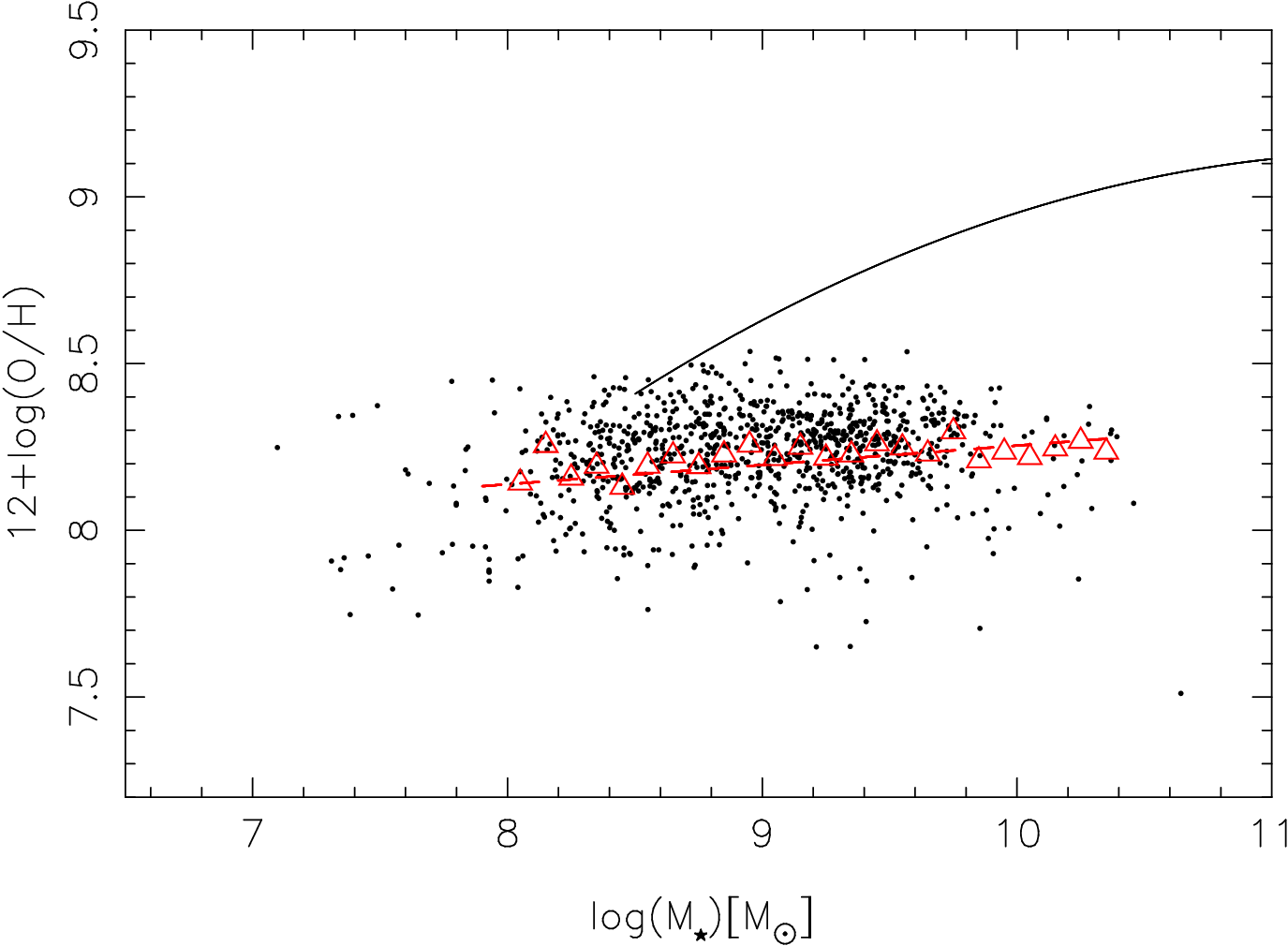}
\end{minipage}
\begin{minipage}{0.5\textwidth}
\centering
\includegraphics[angle=-90,width=70mm]{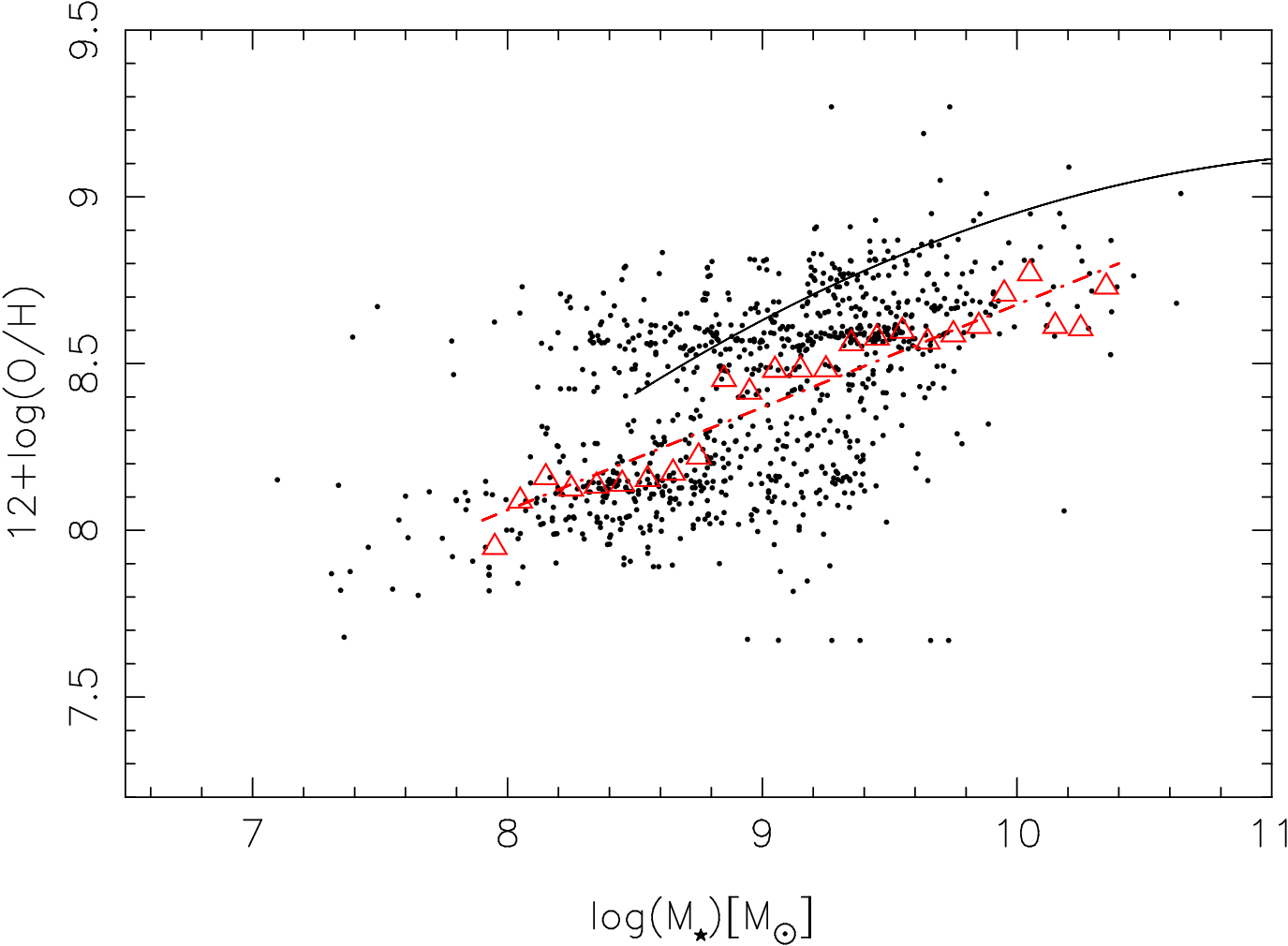}
\end{minipage}
\caption{\baselineskip 3.6mm {T}he 12 + log(O/H) versus $\log
(M_*)$
  relationship. The corresponding stellar mass{es} are the median value{s}
  from SDSS. The dots show 947 star-forming galaxies from our
  work. The triangles show the median from the data in each bin. The
  dashed and dot-dashed lines show the fitted results with the median
  data, respectively. {\it Left}: oxygen abundances are calculated with
  {the }$T_{\rm e}$ method; {\it Right}:
   oxygen abundances are calculated with the $R_{23}$
  method.}
\label{fig.oh.com}
\end{figure*}

When we try to derive the relations of stellar masses and
metallicities of galaxies, we would like to take the galaxies
having $T_{\rm e}$ estimates and $T_{\rm e}$-based O/H. However,
the question is: Can the{ galaxies selected with} [O~{\sc
iii}]$\lambda$4363 be good representatives of the whole{
population of} galaxies? To understand this, we put all our sample
galaxies in the MZR and compare with the main sample{ of SDSS}
galaxies. These are given in Figure~\ref{fig.oh.com}. Both the
abundances from {the }$T_{\rm e}$-method (the left panel) and
strong-line method (the right panel) are compared with Tremonti's
results (the solid line) from the SDSS main galaxy sample in{
terms of} MZR. In both panels, the discrepancies between the data
points and the solid line are quite clear, meaning these objects
could not be {ideal} representatives of the general{ population
of} galaxies. They could be bias{ed} to{ galaxies with} lower
metallicity.  Comparing the left to the right panels, it is clear
{that an} obvious difference{ exists} between the $T_{\rm
e}$-based{ cases} and the{ oxygen abundances derived with the}
strong{ }line method. The dashed and dot-dashed lines show the{ir}
median trends, respectively.

\begin{figure}
\centering
\includegraphics[angle=-90,width=8cm]{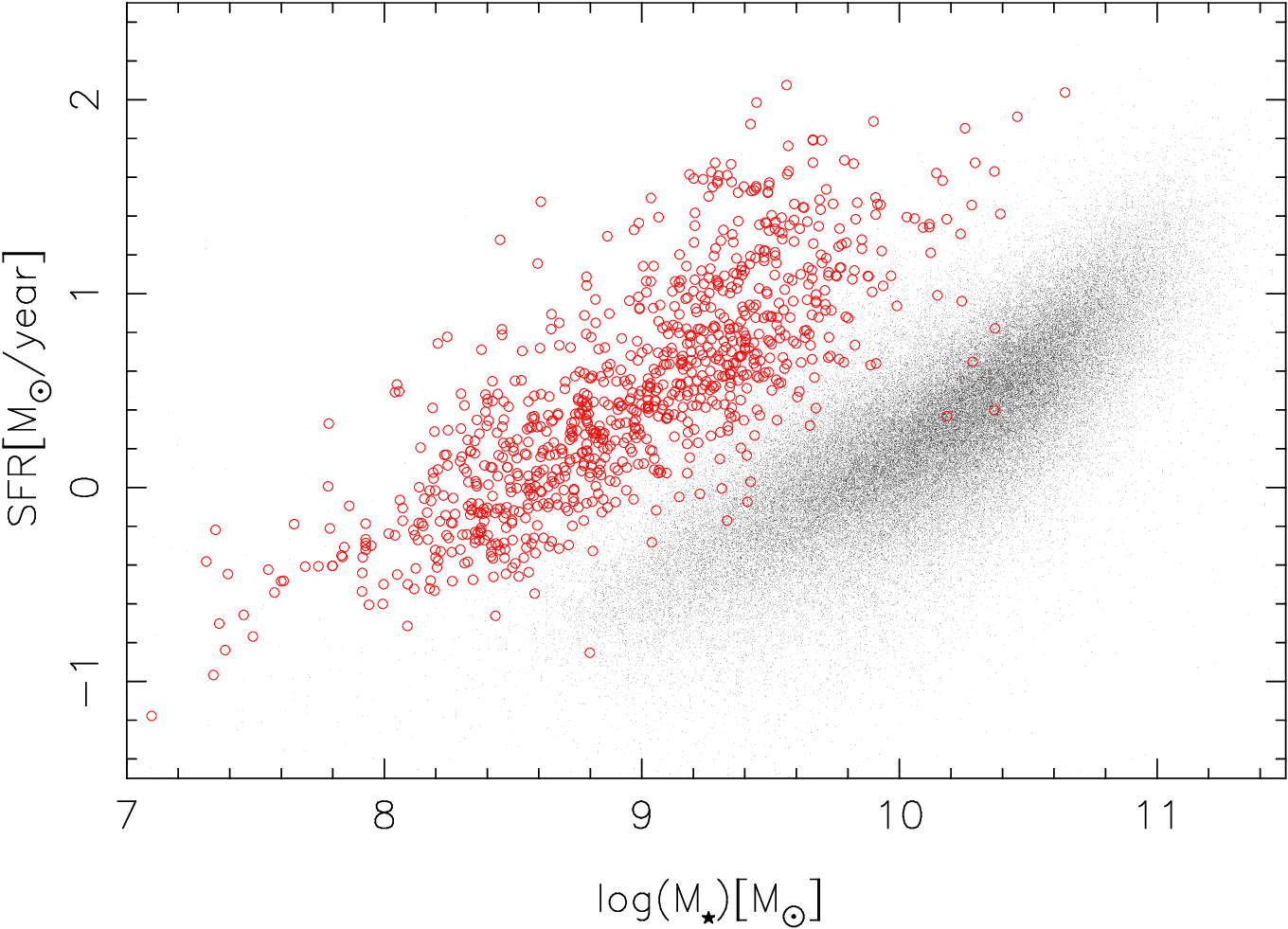}

\caption {\baselineskip 3.6mm The relations of SFR vs. stellar
mass. The circles are the galaxies whose [O~{\sc
iii}]$\lambda$4363 is larger than
  5$\sigma$. The small dots are {those from }the SDSS main galaxy
  sample. }
\label{fig.sfr-mass}
\end{figure}

To understand the real reason {for} such{ a} discrepancy and to
check how {probable it is that} galaxies with{ detected} [O~{\sc
iii}]$\lambda$4363 represent the properties of {a }general{
population of} galaxies, we plot the relations of SFR vs. stellar
mass for galaxies in our sample and the ones in{ the} SDSS main
galaxy sample. This is given in Figure~\ref{fig.sfr-mass}.
 The discrepancy is quite clear.  The galaxies in our
sample are{ more} bias{ed} to the ones having higher
SFRs than the normal galaxies at {a} given stellar mass.  This can
be understood since these low-metallicity galaxies should have
strong emission lines and higher SFRs.

Therefore, we should be very careful if we want to use the galaxies
having $T_{\rm e}$([O~{\sc iii}]) estimates and $T_{\rm e}$-based oxygen
abundances to derive{ the} MZR of galaxies, since maybe they are not the
typical representatives of the general{ population of} galaxies.

{H}ere we may agree with \cite{sta05}: for the metal-rich (massive)
ones, using [O~{\sc iii}]$\lambda$4363 leads to a strong
selection bias towards{ galaxies with} low metal abundance.

Shown {in} Figure{s}~\ref{fig.oh.com} and~\ref{fig.sfr-mass}, the
selected galaxies with reliable [O~{\sc iii}]$\lambda$4363
measurements are bias{ed} to low metallicities.  This can be
understood from the basic physics of photoionized nebulae (Garnett
1992; Stasinska 2002; Ferland 2003). [O~{\sc iii}] is usually the
most efficient coolant in ionized nebulae. In the HII region of
metal-poor galaxies, {there are} few {cooling ions }in the
interstellar{ medium; because} the temperature of plasma is high,
more $\rm O^{++}$ ions stay in higher energy states, {so} the
[O~{\sc iii}]$\lambda$4363 is easily measured. At high
metallicity, cooling is efficient since the heavy elements enforce
the cooling effect{;} the temperature is low, the collision
excitation to higher energy state{s is greatly} reduce{d},{ and}
therefore the temperature-sensitive emission lines are very
difficult to detect. {T}he emission lines with lower excitation
potential could be detected more easily, such as{ the strong
emission lines of} [O~{\sc iii}]$\lambda\lambda$4959,5007.

\subsection{Re-deriv{ing} the Mass-metallicity Relations {with} a
Wide{ Range of} Stellar Mass from $10^6~M_{\odot}$ to
$10^{11}~M_{\odot}$ }

\begin{figure}[h!!]

\vs
 \centering
\includegraphics[angle=-90,width=8.9cm]{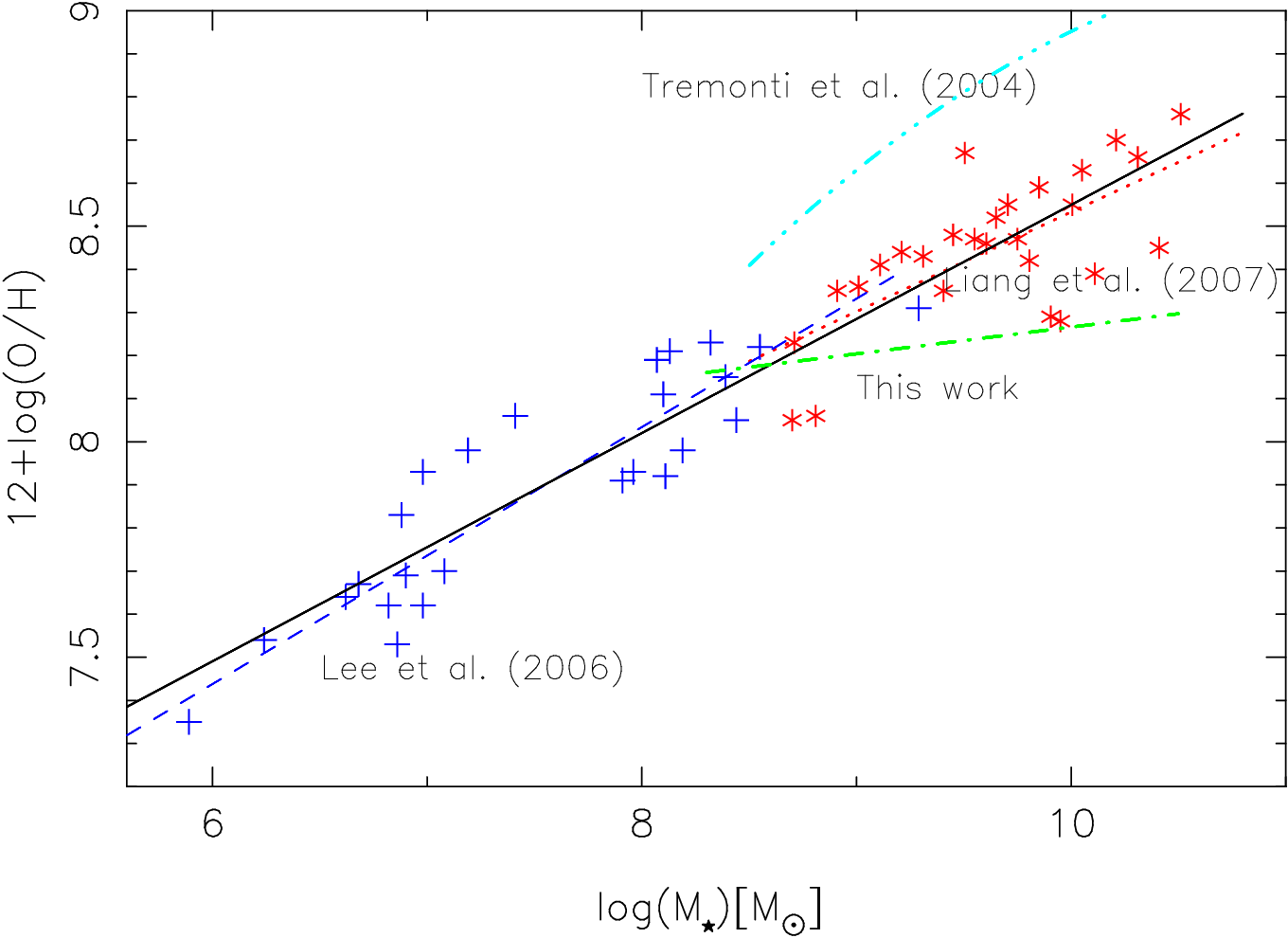}

\vspace{-2.5mm}
 \caption {\baselineskip 3.6mm Fit {of }the
mass-metallicity relations with \cite{lee06} and
  \cite{lia07} data in a wide range of stellar mass from
  10$^6~M_{\odot}$ to 10$^{11}~M_{\odot}$. The data{ follow a straight line} well (the black solid line). This shows the
  {MZR}s of galaxies whose oxygen abundance{ is}
  derived from {the }$T_{\rm e}$ method. The dotted line show{s} the one from{ the}
  \cite{lia07} data. \cite{lee06} fit the data to a line which
  {is}
  shown {as} {the } dashed line. The fit from \cite{tre04} is
  the dot-dot-dashed line, and the dot-dashed line shows the fit for
  our sample{ of} galaxies discussed in Sects.~2 and 3. }
\label{fig.compare.R23}

\vs\vs\centering
\includegraphics[angle=-90,width=9cm]{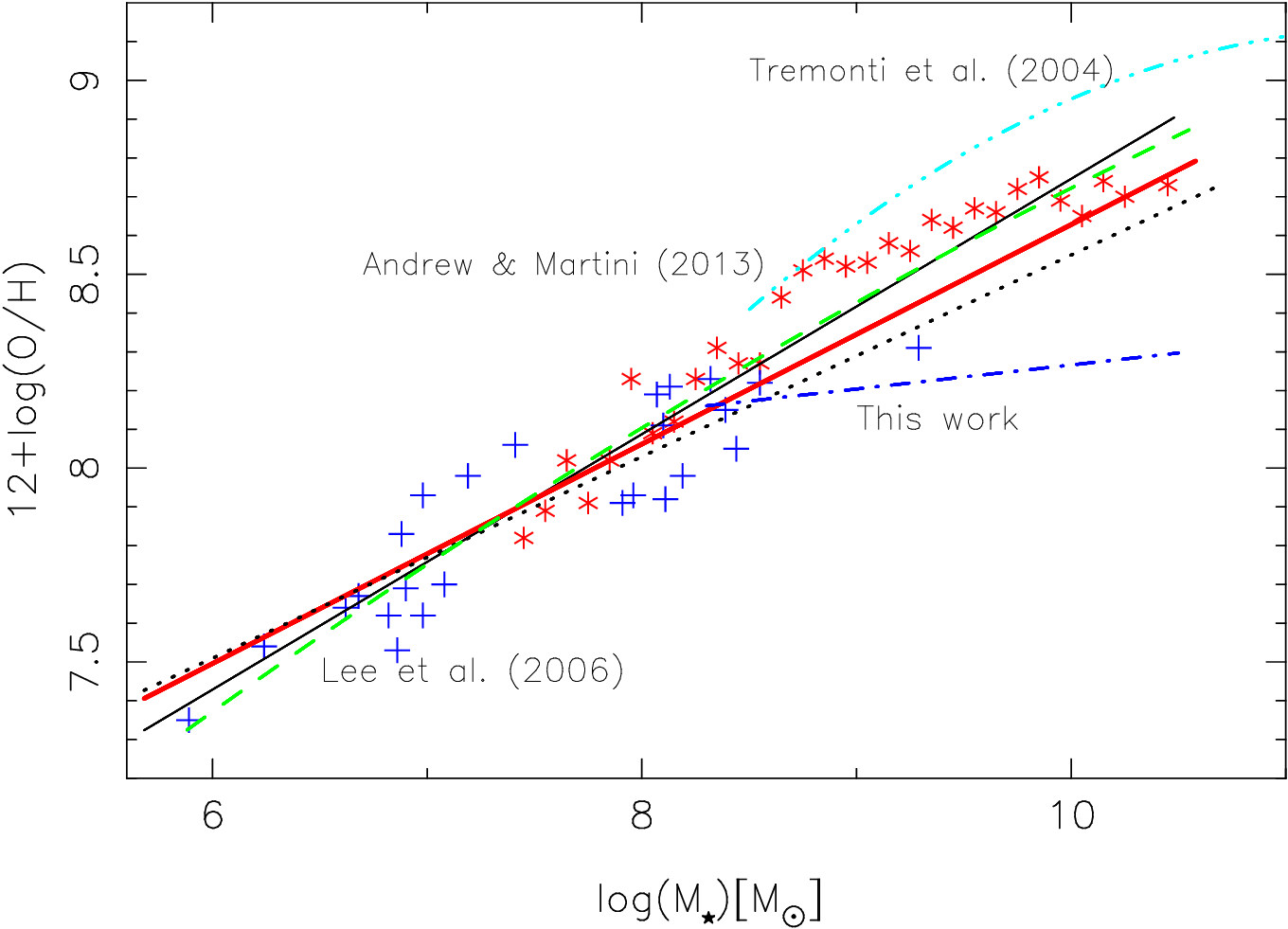}

\vspace{-2mm} \caption {\baselineskip 3.6mm {C}ombin{ing} the data
from \cite{lee06}, \cite{lia07} and \cite{and13} altogether{,}
{we} get their fitted line which is show{n} {as} a thick solid
line. We do a linear fit with{ data from} \cite{lee06} and
\cite{and13}{,}
  show{n} {as} a{ thin} solid line. We also fit{ data from} \cite{lee06} and \cite{and13}
  with a polynomial which {is} plotted {as} a
  dashed{ }line. We display the fitted line{ of data} from \cite{lee06}
   and \cite{lia07} with a dotted line. The
  fits from \cite{tre04} and from our sample{ of} galaxies are
  also show{n} {as} the dot-dot-dashed line and the dot-dashed line,
   respectively.}
\label{Andrews_Te_line}
\end{figure}

It is necessary to re-derive the relationship between (O/H)
abundances and stellar masses for a wide range of stellar mass.
Here we take the sample{ of} galaxies from \cite{lee06} for low
mass galaxies down to 10$^{11}\,M_{\odot}$,{ and} \cite{lia07} and
\cite{and13} for massive galaxies obtained through stacking the
spectra of the multipl{e} galaxies.  \cite{lia07} measured $T_{\rm
e}$[O~{\sc ii}] from the [O~{\sc ii}]$\lambda\lambda$ 7320, 7330
lines and then inferred $T_{\rm e}$[O~{\sc iii}] (and the $\rm
O^{++}$ ionic abundance) from{ analyzing} the $t_2 - t_3$ relation
show{n} {in} our Equation~(\ref{eq3}). \cite{and13} measured both
$T_{\rm e}$[O~{\sc iii}] and $T_{\rm e}$[O~{\sc
  ii}] for some stacked objects. These differences could be partly
responsible for the offset between \cite{and13} and \cite{lia07} in
MZR.

By combining data from \cite{lee06} and those from \cite{lia07} or
\cite{and13}, we can derive the MZR of galaxies with {$T_{\rm
e}$}-based O/H abundances for a wide range of stellar mass{es}
from 10$^6$ to 10$^{11}M_{\odot}$.

The relations derived from the joint sample of \cite{lee06} and
\cite{lia07} are {shown} in Figure~\ref{fig.compare.R23}.
 \cite{lia07} gave an equation $12+
  \log ({\rm O/H}) = 6.223 + 0.231 \times \log M_{*}$ (the dotted line in
Fig.~\ref{fig.compare.R23}) and \cite{lee06} present the equation
$12+\log ({\rm O/H}) = 5.65 + 0.298 \times
  \log M_{*}$ (the dashed line) for their{ respective} samples.
  The new relation of MZR {covering a} wide range
of stellar mass is given as (the solid line in
Fig.~\ref{fig.compare.R23})
\begin{equation}
\label{eq9}
 12+\log({\rm O/H}) =0.260 \log M_{*} + 5.950\, .
\end{equation}
In Figure~\ref{fig.compare.R23}, the median trend of the{ objects detected with} [O~{\sc
  iii}]$\lambda$4363, our working sample, is shown as
the dot-dashed line, and the trend of \cite{tre04} for{ the} SDSS
main galaxy sample is shown as the dot-dot-dashed line.

We also derive the MZR from the joint sample of \cite{lee06} and
\cite{and13} in this wide{ range of} stellar mass. We fit these
data with a least{ }squares linear fit. The result is shown by the
thin
solid line in Figure~\ref{Andrews_Te_line}. 
 The
fitt{ing result} is as follows
\begin{equation}
 12+ \log ({\rm O/H}) =0.329 \log M_{*} + 5.453\,.
\end{equation}
We also fit{ data from} \cite{lee06} and \cite{and13} with a polynomial shown as the dashed line in
Figure~\ref{Andrews_Te_line}. The fitted polynomial is as follows (the dashed line)
\begin{equation}
 12+ \log({\rm O/H}) = -0.01417 (\log M_{*})^2 + 0.56430 \log M_{*} +
 4.49612\, {.}
\end{equation}
The dotted line in Figure~\ref{Andrews_Te_line} is from
Equation~(\ref{eq9}), {which is }the solid one in
Figure~\ref{fig.compare.R23}.  The difference between \cite{lia07}
and \cite{and13} in {the }metal-rich region is more obvious. This
could be due to several reasons, such as \cite{lia07} use {a
smaller} sample, {i.e. }only select{ing} those objects having
higher {equivalent width} ([O~{\sc ii}]){,} which could {be
slightly }bias{ed} to lower metallicity, and the method in
\cite{and13} (IRAF) overestimate{s} the log($\rm O^{++}/\rm H^+$)
by 0.1\,dex compar{ed} with the method used in \cite{lia07}
(Izotov 2006 formula) (see the left panel of
Fig.~\ref{fig.o3.o2}).

To minimize the bias from different samples in different
{studies}, we combine the data from \cite{lee06}, \cite{lia07} and
\cite{and13} altogether and get the least square{s} fit for their
mass-metallicity relations. The fitted line is show{n} {as} a
thick solid line in Figure~\ref{Andrews_Te_line}. For comparison,
we display the fitted line from{ the} \cite{lee06} and
\cite{lia07} data with the dotted line and the line from
\cite{lee06} and \cite{and13} data with the thin solid line. The
MZR fitted with the three data{ sets} is given as
\begin{equation}
\label{offset}
 12+ \log ({\rm O/H}) =0.283 \log M_{*} + 5.798\, .
\end{equation}

\section{summary}

We select 947 star-forming galaxies from SDSS{,} which have
detected [O~{\sc iii}]$\lambda$4363 emission lines. This can help
to reliably estimate their electron temperature $T_{\rm e}$ and
then $T_{\rm e}$-based oxygen abundances. {With} this sample, we
first carefully compare the electron temperature in high
ionization regions{,} $t_3${,} from different methods and {that}
in low ionization regions{, $t_2$,} from different methods. {T}hen
the result{ing} ionic abundances $\rm O^{++}$/H$^+$ and $\rm
O^{+}$/$\rm H^+$ are compared correspondingly. In the second part
of this work, we derive the MZR from the $T_{\rm e}$-based oxygen
abundances of galaxies {from} literature in a wide range of
stellar mass, from 10$^6~M_{\odot}$ to 10$^{11}~M_{\odot}$. We
find that
\begin{itemize}
  \item[(1)] The $t_3$ electron temperature{s} derived from \cite{izo06}
and {\sc temden.nebular} in {\sc stadas.iraf} are quite similar.

\item[(2)] The $t_2$ electron temperatures{ resulting} from relations of
three analyses that considered $t_2$--$t_3$ show obvious
difference{s}, which result in slightly different oxygen
abundances.

\item[(3)] The ionic abundances $\rm O^{++}/\rm H^+$ {are}
higher than $\rm O^{+}/\rm H^+$ for most cases.

\item[(4)] The $T_{\rm e}$-derived and the strong-line derived oxygen abundances show{ a}
clear discrepancy, which is more obvious {with} increasing stellar
mass $\log (M_{*})$.

\item[(5)] The sample{ of} galaxies from SDSS {with} detected [O~{\sc
iii}]$\lambda$4363 have lower metallicit{i}es and higher SFRs.
They may not be representative of the whole{ population of}
galaxies and their properties.

\item[(6)] We gather the galaxies having $T_{\rm e}$ and $T_{\rm
e}$-derived oxygen abundances in {a }wide range of stellar mass
from 10$^6~M_{\odot}$ to 10$^{11}~M_{\odot}$ (Lee et al. (2006)
for low mass galaxies, {and} \cite{lia07} and \cite{and13} for
massive galaxies). Then we derive new relations of stellar masses
and metallicities by combining the data from low mass to massive
ones.
\end{itemize}

In the future, more data should be obtained{,} especially in {the
}metal-poor region; more work should be done to analyze the
mass-metallicity relation in detail bas{ed} {on} the $T_{\rm
e}$-derived oxygen abundances.

\begin{acknowledgements}
  We thank the referee for very helpful comments and suggestions{,} which significantly
  improved this paper. This work was supported by the National Natural
  Science Foundation of China (Grant Nos.~10933001, 11273026, 11390371, 11178013,
  11233004, U1331104 and 11373026), and by the Natural
  Science Foundation of Shan{d}ong {Province }(ZR2010AM006).
\end{acknowledgements}

\end{document}